\documentstyle[12pt]{article}
\begin{document}

\begin{center}
{\Large
SEMICLASSICAL SYMMETRIES
}
\\
{{\large  O.Yu.Shvedov} \\
{\it
Sub-Dept. of Quantum Statistics and Field Theory},\\
{\it Dept. of Physics, Moscow State University},\\
{\it 119899, Moscow, Vorobievy Gory, Russia}
}
\end{center}

\def\qp{
\mathrel{\mathop{\bf x}\limits^2},
\mathrel{\mathop{-i\frac{\partial}{\partial {\bf x}}}\limits^1} 
}

\begin{abstract}

Essential properties  of  semiclassical  approximation   for   quantum
mechanics are viewed as axioms of an abstract semiclassical mechanics.
Its symmetry properties are  discussed.  Semiclassical  systems  being
invariant under Lie groups are considered.  An infinitesimal analog of
group relation is written.  Sufficient conditions  for  reconstructing
semiclassical group  transformations  (integrability of representation
of Lie algebra) are discussed.  The obtained results may be  used  for
mathematical proof  of Poincare invariance of semiclasical Hamiltonian
field theory and for investigation of quantum anomalies.

\end{abstract}
{\sl math-ph/0109016}

\footnotetext{e-mail:  shvedov@qs.phys.msu.su}

\newcounter{eqn}[section]
\renewcommand{\theeqn}{\thesection.\arabic{eqn}}
\def\lab{\refstepcounter{eqn}\eqno(\thesection.\arabic{eqn})}
\def\l#1{\lab\label{#1}}
\def\r#1{(\ref{#1})}
\def\c#1{\cite{#1}}
\def\i#1{\bibitem{#1}}
\def\be{$$}
\def\ee{$$}
\def\bea#1 \eea{$$ \begin{array}{c} #1 \end{array} $$}
\def\bea#1 \l#2 \eea{$$ \begin{array}{c} #1 \end{array} \l{#2} $$}
\def\beb#1 \eeb{$$ \begin{array}{c} #1 \end{array} $$}
\def\crr{\\}
\sloppy

\newpage

\section{Introduction}

Semiclassical approximation is widely used in  quantum  mechanics  and
field theory.  There  are only few cases when the Schrodinger equation
possesses exact solutions.  It is then necessary to develop  different
approximation techniques  in order to investigate evolution equations.
Semiclassical method is universal:  it may be applied,  provided  that
the coefficients  of all derivative operators are small,  of the order
$O(\lambda)$ as  $\lambda\to  0$,  while  an  explicit  form  of   the
Hamiltonian may   be   arbitrary.

There are different semiclassical ansatzes that approxiomately satisfy
quantum mechanical equations. They are reviewed in section 2. The most
popular semiclassical substitution is the WKB-ansatz.  However,  there
are other wave functions (for example,  the Maslov complex-WKB  ansatz
\c{M1,M2}) that  conserve  their forms under time evolution in the
semiclassical approximation.

Semiclassical conception can  be formally applied to quantum field
theory (QFT)  under certain conditions \c{J}.  Examples of application
of semiclassical conceptions are soliton quantization  \c{J,DHN,R,GJ},
quantum  field  theory  in  a  strong  external background \c{GMM,BD},
one-loop \c{B1,B2,B11,B12},  time-dependent  Hartree-Fock  \c{HF1,HF2}
and Gaussian approximations \c{G1,G2,G3,G4}.

Unfortunately, "exact"   quantum   field   theory    is    constructed
mathematically for  a  restricted  class  of  models  only  (see,  for
example, \c{H,GJ,Ar1,Ar2}).  Therefore,  formal approximate methods such
as perturbation  theory  seem  to be ways to quantize the field theory
rather than to construct approximations for  the  exact  solutions  of
quantum field  theory equations.  The conception of field quantization
within the perturbation framework is popular \c{BS,SF}. One can expect
that the semiclassical approximation plays an analogous role.

An important axiom of QFT  is  the  property  of  Poincare  invariance
\c{A2,A3}. There  are  also  other  symmetries  in QFT,  as well as in
different models of quantum mechanics.

The purpose of this paper is to  introduce  a  notion  of  a  symmetry
transformation in   the   semiclassical   mechanics,  as  well  as  to
investigate infinitesimal  properties  of  groups   of   semiclassical
transformations, especially for the case of field systems.

\section{Semiclassical mechanics}

\subsection{Semiclassical substitutions to quantum mechanical equations}

This subsection deals with a review of semiclassical substitutions for
the finite-dimensional equations of the form
\be
i\lambda \frac{\partial \psi_t (x)}{\partial t} = H_t (x,  -  i\lambda
\frac{\partial}{\partial x})  \psi_t(x),  \qquad  t \in {\bf R},  x\in
{\bf R}^d,
\l{1.1}
\ee
where $H_t(q,p)$ is an arbitrary function.

\subsubsection{WKB and Maslov complex-WKB wave functions}

The most  famous  semiclassical  approach is the WKB-method.  It is as
follows. The initial condition for eq.\r{1.1} is chosen to be
\be
\psi_0(x) = \varphi_0(x) e^{\frac{i}{\lambda}S_0(x)},
\l{1.2}
\ee
where $S_0$ is a real function. The WKB-result \c{Maslov1} is that the
solution of eq.\r{1.1} at time moment $t$ has the same type \r{1.2} up
to $O(\lambda)$,
$$
||\psi_t - \varphi_t e^{\frac{i}{\lambda}S_t}|| = O(\lambda).
$$
The Hamilton-Jacobi  equation  for $S_t$ and the transport equation
for $\varphi_t$ can be written \c{Maslov1}.

However, we are not  obliged  to  choose  the  initial  condition  for
eq.\r{1.1} in  a form \r{1.2}.  There are other substitutions to
eq.\r{1.1} that  conserve  their  forms  under   time   evolution   as
$\lambda\to 0$.  For  example,  consider  the  Maslov complex-WKB wave
function \c{M1,M2,MS3},
\be
\psi_0(x) =  const  e^{\frac{i}{\lambda}   S_0}   e^{\frac{i}{\lambda}
P_0(x-Q_0)} f_0(\frac{x-Q_0}{\sqrt{\lambda}})                   \equiv
(K^{\lambda}_{S_0,Q_0,P_0} f_0)(x),
\l{1.3}
\ee
which corresponds to uncertanties of coordinates  and  momenta  of  the
order $O(\sqrt{\lambda})$,  since  the  smooth  function $f_0(\xi)$ is
chosen to damp rapidly at  spatial  infinity.  Quantities  $Q_0,P_0\in
{\bf R}^d$  may  be interpretted as classical values of coordinates and
momenta.

It happens that the initial condition \r{1.3} conserves its form under
time evolution up to $O(\sqrt{\lambda})$ \c{M1,M2},
\be
\psi_t(x) =  const  e^{\frac{i}{\lambda}   S_t}   e^{\frac{i}{\lambda}
P_t(x-Q_t)} f_t(\frac{x-Q_t}{\sqrt{\lambda}}) + O(\sqrt{\lambda}).
\l{1.4}
\ee
Moreover, $Q_t,P_t$   satisfy  the  classical  Hamiltonian  equations,
$S_t-S_0$ is the action on the classical trajectory,  while $f_t(\xi)$
satisfies the Schrodinger equation with a quadratic Hamiltonian.

Semiclassical state \r{1.3} may be interpretted as a point on a bundle
({\it "semiclassical bundle"} \c{Shv1,Shv2}).  The base of the  bundle
is a manifold ${\cal X} = \{ (S_0,Q_0,P_0) | S_0 \in {\bf R},  Q_0,P_0
\in {\bf R}^d \}$ which may be called as an extended phase  space.  If
the point $X_0 = (S_0,Q_0,P_0) \in {\cal X}$ is given, the "classical"
state is specified.  However,  one should also  specify  the  function
$f_0(\xi)$ (the  "shape"  of  the  wave  packet).  This corresponds to
choice of an element of the fibre. The fibres of the bundle are spaces
${\cal S}({\bf  R}^{d})$, so that the bundle is trivial.
If  the  point $(X_0 \in {\cal X},  f_0 \in
{\cal S}({\bf R}^d))$ is given, the semiclassical wave function \r{1.3}
is completely specified.

The semiclassical   evolution  transformation  may  be  viewed  as  an
automorphism of  the  semiclassical  bundle,   since   the   evolution
transformation of   $(S,Q,P)$   does   not   depend   on   $f_0$.  The
transformations $u_t:  {\cal X} \to {\cal X}$  and  unitary  operators
$U_t^0(u_tX\gets X): f_0 \mapsto f_t$ are then given. One can consider
the completion ${\cal F} = L^2({\bf R}^d)$ of the space ${\cal S}({\bf
R}^d)$ and extend the unitary operators $U_t^0(u_tX \gets X)$ to $\cal
F$. One obtain then operators $U_t(u_tX \gets X) :  {\cal F} \to {\cal
F}$.

\subsubsection{Maslov theory of Lagrangian manifolds with complex germ}

The wave function  \r{1.2}  rapidly  oscillates  with  respect  to  all
variables. The  wave  function  \r{1.3}  rapidly  damps  at  $x-Q_t  >>
O(\sqrt{\lambda})$. One should come to the conclusion
that  there  exists  a  wave
function asymptotically  satisfying  eq.\r{1.1}  which  oscillates with
respect to one group of variables and  damps  with  respect  to  other
variables. The  construction  of  such  states  is given in the Maslov
theory of  Lagrangian  manifolds  with  comples  germ  \c{M1,M2}.  Let
$\alpha \in {\bf R}^k$,  $(P(\alpha),Q(\alpha)) \in {\bf R}^{2d}$ be a
$k$-dimensional surface   in   the   $2d$-dimensional   phase   space,
$S(\alpha)$ be a real function, $f(\alpha,\xi)$, $\xi\in {\bf R}^d$ is
a smooth function.  Set $\psi(x)$ to be not exponentially small if and
only if  the  distance between point $x$ and surface $Q(\alpha)$ is of
the order $\le O(\sqrt{\lambda})$.  Otherwise, set $\psi(x) \simeq
0$. If   $\min_{\alpha}   |x-Q(\alpha)|  =  |x-Q(\overline{\alpha})|  =
O(\sqrt{\lambda})$, set
\be
\psi(x) =                                              c_{\lambda}
e^{\frac{i}{{\lambda}}S(\overline{\alpha})}
e^{\frac{i}{{\lambda}}                        P(\overline{\alpha})
(x-Q(\overline{\alpha}))}
f(\overline{\alpha}, \frac{x-Q(\overline{\alpha})}{\sqrt{\lambda}}).
\l{w4}
\ee

One can note that wave functions \r{1.2} and \r{1.3} are  partial  cases
of the   wave   function   \r{w4}.  Namely,  for  $k=0$  the  manifold
$(P(\overline{\alpha}), Q(\overline{\alpha}))$ is a point,  so that the
functions \r{w4} coincide with \r{1.3}.
Let $k=d$.      If      the      surface       $(P(\overline{\alpha}),
Q(\overline{\alpha}))$ is  in  the  general  position,  for $x$ in some
domain one has $x= Q(\overline{\alpha})$ for some $\overline{\alpha}$.
Therefore,
$$
\psi(x) =          c_{\lambda}          e^{\frac{i}{{\lambda}}
S(\overline{\alpha})}  f(\overline{\alpha},0).
$$
We obtain the WKB-wave function.  Thus, WKB and wave-packet asymptotic
formulas \r{1.2}  and  \r{1.3}  are  partial  cases of the wave function
\r{w4} appeared in the theory of  Lagrangian  manifolds  with  complex
germ.

The lack    of    formula   \r{w4}   is   that   the   dependence   of
$\overline{\alpha}$ on $x$ is implicit and too  complicated.  However,
under certain    conditions    formula    \r{w4}   is   invariant   if
$\overline{\alpha}$ is  shifted   by   a   quantity   of   the   order
$O(\sqrt{\lambda})$. In  this case,  the point $\overline{\alpha}$
can be chosen in arbitrary way such  that  the  distance  of  $x$  and
$Q(\overline{\alpha})$ is of the order $O(\sqrt{\lambda})$.

Namely,
\bea
e^{\frac{i}{{\varepsilon}} S(\overline{\alpha}  +   \sqrt{\varepsilon}
\beta)}
e^{\frac{i}{{\varepsilon}}
P(\overline{\alpha}  +   \sqrt{\varepsilon} \beta)
(x - Q(\overline{\alpha}  +   \sqrt{\varepsilon} \beta)) }
f(\overline{\alpha}  +   \sqrt{\varepsilon} \beta,
\frac{x-
Q(\overline{\alpha}  +   \sqrt{\varepsilon} \beta)
}{\sqrt{\varepsilon}}
) \crr
\simeq
e^{\frac{i}{{\varepsilon}} S(\overline{\alpha})}
e^{\frac{i}{{\varepsilon}}
P(\overline{\alpha})
(x - Q(\overline{\alpha})) }
f(\overline{\alpha},
\frac{x-
Q(\overline{\alpha})
}{\sqrt{\varepsilon}} )
\l{w5}
\eea
if
\be
\frac{\partial S}{\partial \overline{\alpha}_i} =
P \frac{\partial Q}{\partial \overline{\alpha}_i}
\l{w6}
\ee
\be
e^{i \beta (\xi \frac{\partial P}{\partial \overline{\alpha}}
- \frac{1}{i} \frac{\partial}{\partial \xi}
\frac{\partial Q}{\partial \overline{\alpha}})} f = f
\l{w7}
\ee
To obtain eqs.\r{w6} and \r{w7},  one should expand left-hand side  of
eq.\r{w5}. Considering   rapidly   oscillating   factors,   we  obtain
eq.\r{w6}. To obtain eq.\r{w7}, it is sufficient to consider the limit
${\lambda}\to 0$.

Conditions \r{w6},  \r{w7}  simplify  the  check  \c{M1} that the wave
function \r{w4} approximately satisfies  eq.\r{1.1}  if  the  functions
$S,P,Q,f$ are time-dependent.

\subsubsection{Composed semiclassical states}

The form \r{w4} of the semiclassical state appeared in the  theory  of
Lagrangian manifolds   with   complex   germ  is  not  convenient  for
generalization to systems of infinite number of degrees of freedom. It
is much  more  convenient to consider to consider wave function \r{1.3}
as an ''elementary'' semiclassical state and wave function \r{w4} as a
''composed'' semiclassical  state  presented  as  a  superposition  of
elementary semiclassical states:
\be
\psi(x) = C_{\lambda} \int d\alpha
e^{\frac{i}{{\lambda}} S(\alpha)}
e^{\frac{i}{{\lambda}}P(\alpha) (x-Q(\alpha))}
g(\alpha, \frac{x-Q(\alpha)}{\sqrt{\lambda}}),
\l{w8}
\ee
where $g(\alpha,\xi)$  is  a  rapidly  damping  function  as   $\xi\to
\infty$. Superpositions  of  such  type were considered in \c{P,K,KV};
the general  case  was  investigated  in  \c{MS3,MS4}.  The   composed
semiclassical states   for  the  abstract  semiclassical  theory  were
studied in \c{Shv2}.

To show that expression \r{w8} is in agreement  with  formula  \r{w4},
notice that  the  wave  function  \r{w8} is exponentially small if the
distance between  $x$  and  the  surface  $Q(\alpha)$  is   of   order
$>O(\sqrt{\lambda})$. Let     $\min_{\alpha}     |x-Q(\alpha)|    =
O(\sqrt{\lambda})$ and         $|x-Q(\overline{\alpha})|         =
O(\sqrt{\lambda})$. Consider    the    substitution    $\alpha   =
\overline{\alpha} + \sqrt{\lambda} \beta$. We find
$$
\psi(x) = C_{\lambda} {\lambda}^{k/2} \int d\beta
e^{\frac{i}{{\lambda }}
S(\overline{\alpha}  +   \beta \sqrt{\lambda})}
e^{\frac{i}{{\lambda}}
P(\overline{\alpha}  +   \beta \sqrt{\lambda})
(x- Q(\overline{\alpha}  +   \beta \sqrt{\lambda}))
}
g(\overline{\alpha}  +   \beta \sqrt{\lambda},
\frac{x- Q(\overline{\alpha}  +  \beta
\sqrt{\lambda})}{\sqrt{\lambda}}
)
$$
If the condition \r{w6} is not satisfied,  this is an  integral  of  a
rapidly oscillating   function.   It  is  exponentially  small.  Under
condition \r{w6} one can consider a  limit  ${\lambda}\to  0$  and
obtain the expression \r{w4}, provided that
$$
c_{\lambda}  = C_{\lambda} {\lambda}^{k/2}
$$
and
\bea
f(\overline{\alpha}, \xi) = \int d\beta
e^{i\beta_s (\frac{\partial P_m}{\partial\overline{\alpha}_s} \xi_m  -
\frac{\partial Q_m}{\partial \overline{\alpha}_s} \frac{1}{i}
\frac{\partial}{\partial \xi_m})} g(\overline{\alpha},\xi)
= \crr
(2\pi)^k \prod_{s=1}^k
\delta(\frac{\partial P_m}{\partial \overline{\alpha}_s} \xi_m -
\frac{\partial Q_m}{\partial      \overline{\alpha}_s}     \frac{1}{i}
\frac{\partial}{\partial \xi_m}) g(\overline{\alpha},\xi).
\l{w9}
\eea
Integral representation \r{w8} simplifies  substitution  of  the  wave
function to   eq.\r{1.1}   and  estimation  of  accuracy.

It follows from eq.\r{w9} that the function $f$ is invariant under the
following change of the function $g$ ("gauge transformation"):
\be
g(\alpha,\xi) \to g(\alpha,\xi) +
(\frac{\partial P_m}{\partial   \alpha_s}   \xi_m   -   \frac{\partial
Q_m}{\partial \alpha_s} \frac{1}{i} \frac{\partial}{\partial \xi_m})
\chi_s(\alpha,\xi).
\l{1.10}
\ee
Thus, the  semiclassical  state  is  specified  at  fixed $S(\alpha)$,
$P(\alpha)$, $Q(\alpha)$ not by the function $g$ but by the  class  of
equivalence of functions $g$: two functions are equivalent if they are
related by the transformation \r{1.10}.

This fact can be also illustrated if we  evaluate  the  inner  product
$||\psi||^2$ as ${\lambda}\to 0$:
\beb
||\psi||^2 = C_{\lambda}^2 \int d\alpha d\gamma \int dx
e^{-\frac{i}{{\lambda}} S(\alpha)}
e^{-\frac{i}{{\lambda}}P(\alpha) (x-Q(\alpha))}
g^*(\alpha, \frac{x-Q(\alpha)}{\sqrt{\lambda}})
\crr
e^{\frac{i}{{\lambda}} S(\gamma)}
e^{\frac{i}{{\lambda}}P(\gamma) (x-Q(\gamma))}
g(\gamma, \frac{x-Q(\gamma)}{\sqrt{\lambda}}),
\eeb
The integral  over  $x$ is not exponentially small if $\alpha-\gamma =
O(\sqrt{\lambda})$. After substitution $\gamma =  \alpha  +  \beta
\sqrt{\lambda}$, $x-Q(\alpha) = \xi\sqrt{{\lambda}}$
and considering the limit ${\lambda}\to 0$, we find
\be
||\psi||^2 \simeq C_{\lambda}^2 {\lambda}^{\frac{k+n}{2}} \int
d\alpha (g(\alpha,\cdot), \prod_{s=1}^k 2\pi \delta
(\frac{\partial P_m}{\partial   \alpha_s}   \xi_m   -   \frac{\partial
Q_m}{\partial \alpha_s} \frac{1}{i} \frac{\partial}{\partial \xi_s}
) g(\alpha,\cdot)).
\l{1.11}
\ee

The $k$-dimensional   surface    $\{(S(\alpha),P(\alpha),Q(\alpha))\}$
("isotropic manifols")  in  the extended phase space has the following
physical meaning.  Consider  the  average  value  of  a  semiclassical
observable $A(x,-i{\lambda}\partial/\partial        x)$.        As
${\lambda}\to 0$, one has
\beb
(\psi, A(x, - i{\lambda} \partial/\partial x) \psi) \simeq
C_{\lambda}^2 {\lambda}^{\frac{k+n}{2}} \int
d\alpha A(Q(\alpha),P(\alpha))
(g(\alpha,\cdot), \crr
\prod_{s=1}^k 2\pi \delta
(\frac{\partial P_m}{\partial   \alpha_s}   \xi_m   -   \frac{\partial
Q_m}{\partial \alpha_s} \frac{1}{i} \frac{\partial}{\partial \xi_s}
) g(\alpha,\cdot)).
\eeb
We see that only values of the corresponding classical  observable  on
the surface $\{(Q(\alpha), P(\alpha))\}$ are relevant for calculations
fo average  values  as  ${\lambda}\to  0$.  This  means  that  the
Blokhintsev-Wigner density   function  (Weyl  symbol  of  the  density
matrix) corresponding  to  the   composed   semiclassical   state   is
proportional to the delta function on the manifold
$\{(Q(\alpha), P(\alpha))\}$.

Therefore, elementary semiclassical states  describe  evolution  of  a
point particle,   while   composed   semiclassical  states  (including
WKB-states) describe evolution  of  the  more  complicated  objects  -
isotropic manifolds.

\subsection{Abstract semiclassical mechanics}

Formally, the semiclassical conception can be applied to quantum field
theory \c{MS3,MS-FT}.  A semiclassical complex-WKB state is  specified
by a  set  $X\in  {\cal X}$ of classical variables (real quantity $S$,
field configuration $\Phi({\bf x})$,  canonically conjugated  momentum
$\Pi({\bf x})$)  and a functional $f[\phi(\cdot)]$ (a  "quantum
state in the external field $X$"),  if the functional
Schrodinger representation of the canonical commutation  relations  is
used. However,  usage of this representation seems to be not rigorous.
On the other hand, one can expect that it is possible to specify
a semiclassical complex-WKB state by an element $f$ of a some  (maybe,
$X$-dependent) Hilbert  space  ${\cal  F}_X$ instead of the functional
$f[\phi(\cdot)]$. The structure of a semiclassical bundle remains then
valid for field theory.

Let us  formulate  a definition of a semiclassical system.  One should
write down a list of essential properties ("axioms") of  such  systems
(cf. \c{Shv1,Shv2}). One of them is as follows.

{\bf A1.} {\it
A locally trivial vector bundle $\pi: {\cal Z} \to {\cal X}$ called as
a semiclassical  bundle is specified.  The base of the bundle $\cal X$
("extended phase space") is  a  smooth  (maybe,  infinite-dimensional)
manifold, while  fibres  ${\cal  F}_X$,  $X  \in {\cal X}$ are Hilbert
spaces.
}

Since the  bundle is locally trivial,  one can suppose without loss of
generality that all spaces ${\cal F}_X$  coincide  in  a  sufficiently
small vicinity of each point.

Elementary ("complex-WKB")  semiclassical  states are viewed as points
on the semiclassical bundle.  Composed semiclassical states should  be
viewed as  smooth  mappings $\alpha \in {\Lambda}^k \mapsto (X(\alpha)
\in {\cal X},  g(\alpha) \in {\cal F}_{X(\alpha)}$ for $k$-dimensional
manifolds $\Lambda^k$  with given measure.  However,  to introduce the
inner product like \r{1.11},  it is not sufficient to  use  axiom  A1.
therefore, additional  structures  on  the  semiclassical  bundle  are
necessary.

\subsubsection{Structures on the semiclassical bundle}

An important feature of the finite-dimensional complex-WKB  theory  is
that a $\lambda$-dependent quantum state $K_X^{\lambda} f$ is assigned
to each set $(X \in {\cal X}, f\in {\cal F}_X)$. Moreover, the mapping
$K^{\lambda}_X$ \r{1.3} satisfies the following properties:
\be
(K^{\lambda}_X f, K^{\lambda}_X f) \to_{\lambda\to 0} (f,f),
\l{1.13a}
\ee
provided that $const = \lambda^{-d/4}$;
\be
i\lambda \frac{\partial}{\partial X_i} K^{\lambda}_X  =  K^{\lambda}_X
[\omega_i - \sqrt{\lambda} \Omega_i + ... ] f,
\l{1.14}
\ee
where
\beb
\omega_i dX_i = P_j dQ_j - dS;\\
(\Omega_i dX_i    f)(\xi)   =   (dP_j   \xi_j   -   dQ_j   \frac{1}{i}
\frac{\partial}{\partial \xi_j}) f(\xi).
\eeb
For the operator \r{1.3},  the terms  $...$  entering  to  eq.\r{1.14}
vanish.

Relation \r{1.14}  is an important property of a semiclassical system.
It allows  us  to  introduce  two   additional   structures   on   the
semiclassical bundle:  the  differential  1-form $\omega_idX_i$ on the
extended phase space $\cal X$ ("action form") and the  operator-valued
differential 1-form $\Omega_i dX_i$.

The commutation relations between operators $\Omega_i$ can be obtained
from eq.\r{1.14}. Namely, apply the commutator
$[i\lambda \frac{\partial}{\partial X_i};
i\lambda \frac{\partial}{\partial X_j}]$ being  zero  to  the  quantum
state $K_X^{\lambda} f$. One finds:
\beb
0 = [i\lambda \frac{\partial}{\partial X_i};
i\lambda \frac{\partial}{\partial X_j}] K_X^{\lambda} f =
K^{\lambda}_X
[ i\lambda \frac{\partial}{\partial X_i} + \omega_i  -  \sqrt{\lambda}
\Omega_i + ...;
i\lambda \frac{\partial}{\partial X_j} + \omega_j  -  \sqrt{\lambda}
\Omega_j +  ...]  f  =  \\
K^{\lambda}_X  (\lambda  [\Omega_i;\Omega_j] +
i\lambda (\frac{\partial \omega_j}{\partial X_i} -
\frac{\partial \omega_i}{\partial X_j}) + ...)f.
\eeb
Thus, the following commutation rules are satisfied:
\be
[\Omega_i; \Omega_j] = i
\left(\frac{\partial \omega_j}{\partial X_i} -
\frac{\partial \omega_i}{\partial X_j}\right).
\l{1.15a}
\ee
It is much more convenient  to  present  relations  \r{1.15a}  in  the
exponential form,
$$
\exp(i\Omega_j\alpha_j)
\exp(i\Omega_j\beta_j) =
\exp(i\Omega_j(\alpha_j+\beta_j))
\exp(\frac{i}{2} \alpha_i\beta_j
\left(\frac{\partial \omega_j}{\partial X_i} -
\frac{\partial \omega_i}{\partial X_j}\right))
$$
We come to the following requirement.

{\bf A2.} {\it
A differential  1-form  $\omega$  and  an operator-valued differential
1-form $\Omega$ are specified on $\cal X$:  for each $X\in  {\cal  X}$
and $\delta  X \in T_X{\cal X}$ the real quantity $\omega_X[\delta X]$
and the self-adjoint operator $\Omega_X[\delta X]$ in the space ${\cal
F}_X$ with a dense domain are given. The commutation relation
$$
e^{i\Omega_X[\delta X_1]}
e^{i\Omega_X[\delta X_2]}
=
e^{i\Omega_X[\delta X_1+ \delta X_2]}
e^{\frac{i}{2} d\omega[\delta X_1; \delta X_2]}
$$
is satisfied.
}

\subsubsection{Composed states}

{\bf 1.} To investigate the inner  product  of  composed  states  like
\r{w8} in  the  abstract semiclassical mechanics,  it is convenient to
simplify the expression
\be
K^{\lambda}_{X(\alpha +  \sqrt{\lambda}  \beta)}   g(\alpha),   \qquad
\alpha = (\alpha_1,...,\alpha_k)
\l{1.16}
\ee
as $\lambda \to 0$.  Let us look for a simplification in the following
form:
\be
K^{\lambda}_{X(\alpha)} \tilde{V}_{\lambda}[\alpha,\beta] g(\alpha)
\l{1.17}
\ee
for some operator  $\tilde{V}_{\lambda}[\alpha,\beta]$.  Applying  the
operators $i\frac{\partial}{\partial \beta_a}$ to expressions \r{1.16}
and \r{1.17}, one finds
\beb
K^{\lambda}_{X(\alpha+ \sqrt{\lambda}\beta)} \left[
\frac{1}{\sqrt{\lambda}}
\omega_i (X(\alpha+ \sqrt{\lambda}\beta))
\frac{\partial X_i(\alpha+ \sqrt{\lambda}\beta)}{\partial\alpha_a}
-
\Omega_i (X(\alpha+ \sqrt{\lambda}\beta))
\frac{\partial X_i(\alpha+ \sqrt{\lambda}\beta)}{\partial\alpha_a}
+ ...
\right] g(\alpha) =
\\
K^{\lambda}_{X(\alpha)} i      \frac{\partial}{\partial       \beta_a}
\tilde{V}_{\lambda}[\alpha,\beta] g(\alpha),
\eeb
so that
$$
i      \frac{\partial}{\partial       \beta_a}
\tilde{V}_{\lambda}[\alpha,\beta]
=
\tilde{V}_{\lambda}[\alpha,\beta]
\left[
\frac{1}{\sqrt{\lambda}}
\omega_i (X(\alpha+ \sqrt{\lambda}\beta))
\frac{\partial X_i(\alpha+ \sqrt{\lambda}\beta)}{\partial\alpha_a}
-
\Omega_i (X(\alpha+ \sqrt{\lambda}\beta))
\frac{\partial X_i(\alpha+ \sqrt{\lambda}\beta)}{\partial\alpha_a}
+ ...
\right]
$$
After extracting a c-number factor
$$
\tilde{V}_{\lambda}[\alpha,\beta] = \exp \left[
- \frac{i}{\sqrt{\lambda}}   \omega_i   \frac{\partial   X_i}{\partial
\alpha_a} \beta_a \right] V_{\lambda} [\alpha,\beta],
$$
one finds in the leading order in $\lambda$ the following equation on
$V_{\lambda} [\alpha,\beta]$:
\be
i \frac{\partial}{\partial \beta_a} V_{\lambda}[\alpha,\beta] =
V_{\lambda}[\alpha,\beta]
\left[
\frac{\partial}{\partial\alpha_b} (\omega_i             \frac{\partial
X_i}{\partial \alpha_a})\beta_b     -     \Omega_i      \frac{\partial
X_i}{\partial \alpha_a}.
\right]
\l{1.18}
\ee
Let us look for the solution of this equation in the following form:
$$
V_{\lambda}[\alpha,\beta] =     c(\alpha,\beta)     \exp    [i\Omega_j
\frac{\partial X_j}{\partial \alpha_a} \beta_a]
$$
for some $c$-number factor $c$.  Making use of  commutation  relations
\r{1.15a}, one takes eq.\r{1.18} to the form:
$$
i\frac{\partial \log  c}{\partial\beta_a}  = \omega_j \frac{\partial^2
X_j}{\partial \alpha_a   \partial\alpha_c}   \beta_c   +   \frac{1}{2}
\left(
\frac{\partial \omega_i}{\partial X_j} +
\frac{\partial \omega_j}{\partial X_i}
\right)
\frac{\partial X_i}{\partial \alpha_c}
\frac{\partial X_j}{\partial \alpha_a}\beta_c,
$$
so that one obtains:
\be
K^{\lambda}_{X(\alpha+ \sqrt{\lambda}\beta)} g(\alpha) \sim
e^{-\frac{i}{\sqrt{\lambda}} \omega_j   \frac{\partial   X_j}{\partial
\alpha_a} \beta_a         -         \frac{i}{2}         \beta_a\beta_c
\frac{\partial}{\partial\alpha_c} (\omega_j             \frac{\partial
X_j}{\partial\alpha_a}) }
K^{\lambda}_{X(\alpha)} e^{i\Omega_j    \frac{\partial   X_j}{\partial
\alpha_a} \beta_a} g(\alpha).
\l{1.19}
\ee

{\bf 2.}
Consider now the composed state
$$
\left(
\begin{array}{c}
X(\cdot) \\
g(\cdot)
\end{array}
\right)
\equiv
C_\lambda \int d\alpha K^{\lambda}_{X(\alpha)} g(\alpha),
\qquad
\alpha = (\alpha_1,...,\alpha_k).
$$

Analogously to eq.\r{1.11}, we obtain the following inner product:
\be
<
\left(
\begin{array}{c}
X(\cdot) \\
g(\cdot)
\end{array}
\right)
,
\left(
\begin{array}{c}
X(\cdot) \\
g(\cdot)
\end{array}
\right)
>
= |C_{\lambda}|^2 \lambda^{k/2}
\int d\alpha d\beta (K^{\lambda}_{X(\alpha)} g(\alpha),
K^{\lambda}_{X(\alpha+\sqrt{\lambda}\beta)}
g(\alpha+\sqrt{\lambda}\beta) ).
\l{1.20}
\ee
Consider the  formal  limit  $\lambda\to  0$.  Let  us  make  use   of
eq.\r{1.19}. One should require the isotropic condition
\be
\omega_j \frac{\partial X_j}{\partial \alpha_a} = 0
\l{1.21}
\ee
to be satisfied.  Otherwise,  the integral \r{1.20}  would  contain  a
rapidly oscillating factor and be therefore exponentially small. Under
condition \r{1.21} the inner product \r{1.20} takes the form
\be
<
\left(
\begin{array}{c}
X(\cdot) \\
g(\cdot)
\end{array}
\right)
,
\left(
\begin{array}{c}
X(\cdot) \\
g(\cdot)
\end{array}
\right)>
= \int d\alpha (g(\alpha), \int d\beta
e^{i\Omega_j \frac{\partial X_j}{\partial\alpha_a} \beta_a} g(\alpha)).
\l{1.22}
\ee
The corresponding inner product  space  of  composed  states  requires
additional investigations: see section 5 for details.

Note that  the  derivation  of  formulas \r{1.21},  \r{1.22} is rather
heuristic. More details are presented in \c{Shv1,Shv2}.

\subsubsection{Symmetry transformations}

An evolution transformation  is  viewed  as  an  automorphism  of  the
semiclassical bundle. If the initial elementary semiclassical state is
presented as
$$
K^{\lambda}_{X_0} f_0,
$$
the semiclassical state at time moment $t$ should be presented  in  an
analogous form
$$
K^{\lambda}_{X_t} f_t
$$
as $\lambda  \to  0$.  The  parameters $X_t \in {\cal X}$ are uniquely
specified by the initial values  $X_0  \in  {\cal  X}$,  so  that  the
transformation $u_t: X_0 \mapsto X_t$ should be given. The vector $f_t
\in {\cal  F}_{X_t}$  should  linearly  depend  on  $f_0   \in   {\cal
F}_{X_0}$; moreover,  the  corresponding dependence should be unitary.
Therefore, the unitary operators $U_t(X_t \gets X_0):  f_0  \in  {\cal
F}_{X_0} \mapsto f_t \in {\cal F}_{X_t}$ should be also specified.

One can also expect that evolution $K^{\lambda}_{X_0} f_0 \mapsto
K^{\lambda}_{X_t} f_t$ remains valid if $X_0$  is  $\lambda$-dependent
in a  way  $X_0 = X_0(\alpha+ \sqrt{\lambda} \beta)$.  It follows from
eq.\r{1.19} that the state
$$
e^{-\frac{i}{\sqrt{\lambda}} \omega_j   \frac{\partial   X_{0,j}}{\partial
\alpha_a} \beta_a         -         \frac{i}{2}         \beta_a\beta_c
\frac{\partial}{\partial\alpha_c} (\omega_j             \frac{\partial
X_{0,j}}{\partial\alpha_a}) }
K^{\lambda}_{X_0(\alpha)} e^{i\Omega_j    \frac{\partial   X_{0,j}}
{\partial \alpha_a} \beta_a} f_0(\alpha).
$$
is taken to
$$
e^{-\frac{i}{\sqrt{\lambda}} \omega_j   \frac{\partial   X_{t,j}}{\partial
\alpha_a} \beta_a         -         \frac{i}{2}         \beta_a\beta_c
\frac{\partial}{\partial\alpha_c} (\omega_j             \frac{\partial
X_{t,j}}{\partial\alpha_a}) }
K^{\lambda}_{X_t(\alpha)} e^{i\Omega_j    \frac{\partial   X_{t,j}}
{\partial \alpha_a} \beta_a} f_t(\alpha).
$$
therefore, the forms $\omega$ and $\Omega$ should satisfy the relations
\beb
\omega_i(X_0) \frac{\partial X_{0,i}}{\partial\alpha_a}
= \omega_i(X_t) \frac{\partial X_{t,i}}{\partial\alpha_a}; \\
e^{i\Omega_i(X_t) \frac{\partial X_{t,i}}{\partial\alpha_a} \beta_a}
U_t(X_t\gets X_0) = U_t(X_t\gets X_0)
e^{i \Omega_i(X_0) \frac{\partial X_{0,i}}{\partial\alpha_a}\beta_a}.
\eeb
We come therefore to the following definition.

{\bf Definition  2.1.}  {\it  A   symmetry   transformation   of   the
semiclassical bundle  is  a  set of a smooth mapping $u:  {\cal X} \to
{\cal X}$ and an unitary operator
$U(uX \gets X): {\cal F}_X \to {\cal F}_{uX}$
such that
\be
\omega_{uX} [u^*(uX\gets X) \delta X] = \omega_X [\delta X]
\l{1.23}
\ee
and
\be
e^{i\Omega_{uX} [u^*(uX\gets X) \delta X]}
U(uX\gets X)  = U(uX\gets X)
e^{i\Omega_{X} [\delta X]},
\l{1.24}
\ee
where $X\in {\cal X}$, $\delta X \in T_X{\cal X}$,
$u^*(uX\gets X):  T_X{\cal  X}  \to  T_{uX}  {\cal  X}$  is an induced
mapping of tangent spaces.
}

Consider the symmetry transformation of the composed state:
$$
\left(
\begin{array}{c}
X(\alpha) \\
g(\alpha)
\end{array}
\right)
\mapsto
\left(
\begin{array}{c}
uX(\alpha) \\
U(uX(\alpha) \gets X(\alpha))
g(\alpha)
\end{array}
\right)
$$
It follows   from   eq.\r{1.23}   that   the    isotropic    condition
\r{1.21}conserves under  time evolution.  Eq.\r{1.24} implies that the
inner product \r{1.22} also conserves under time evolution,  so that a
symmetry transformation of a composed state is also isometric.

Several quantum  mechanical  and  quantum  field systems are invariant
under symmetry groups. Being applied to the semiclassical theory, this
property means the following.

{\bf Definition 2.2.} {\it
A semiclassical bundle is invariant under Lie group $\cal G$ if:\\
(i) for   each  $g\in  {\cal  G}$  a  symmetry  transformation  $(u_g;
U_g(u_gX\gets X))$ of the semiclassical bundle is specified;\\
(ii) the mapping $(g,X) \mapsto u_gX$ is smooth and the group property
\be
u_{g_1g_2} = u_{g_1} u_{g_2}
\l{1.5a}
\ee
is satisfied;
\\
(iii) the following group property takes place:
\be
U_{g_1} (u_{g_1g_2}X \gets u_{g_2}X)
U_{g_2} (u_{g_2}X \gets X) = U_{g_1g_2} (u_{g_1g_2} X \gets X).
\l{x0}
\ee
}

It is well-known that classical symmetries may be not hold in  quantum
field theory:   the   quantum   anomalies   arise   in   the  one-loop
approximation. Thus,  properties \r{1.5a},  \r{x0} should be carefully
checked for each classical symmetry.

A useful   approach  for  constructing  group  representations  is  an
infinitesimal method:  one first constructs a  representation  of  the
corresponding Lie algebra and then integrates the representation.  The
theory of   integrability   of    representattions    is    nontrivial
\c{BR,N,FS1,FS2,FS3}.

One can  expect  that  check  of  properties \r{1.5a},  \r{x0} for the
semiclassical mechanics  can  be  also  performed  in  analogous  way.
Section 3 deals with infinitesimal formulations of property \r{x0} and
conditions of integrability  of  algebra  representations.  Since  the
derivartions of   \c{BR,N,FS1,FS2,FS3}  are  not  convenient  for  our
purposes, another condition of integrability is suggested.

Usually, the infinitesimal generators of semiclassical symmetries  are
quadratic Hamiltonians,  while  operators  $\Omega_i$  are  linear  in
coordinates and momenta. Since the most interesting examples are given
by the  field  theory,  the  case  of  the  Fock  spaces ${\cal F}_X$,
generators being quadratic with respect to creation  and  annihilation
operators and   $\Omega_j$  being  linear  combinations  of  them,  is
considered in  section  4.  The  conditions   of   integrability   are
reformulate. Section   5  deals  with  group  transformations  of  the
composed states. Section 6 contains concluding remarks.

\section{Infinitesimal symmetries}

The purpose of this section is to investigate infinitesimal analogs of
eqs.\r{1.5a}, \r{x0}.

\subsection{From groups to algebras}

{\bf 1.}
By $T_e{\cal G}$ we denote the tangent space to the
Lie group  $\cal G$ at $g=e$.  Let $A\in T_e{\cal G}$,  $g(\tau)$ be a
smooth curve on the group $\cal G$ with the tangent vector
$A$ at the point $g(0)=e$.
Introduce the differenial operator
$\delta[A]$ on the space of differentiable functionals
$F$  on $\cal   X$:
\be
(\delta[A]F)(X) = \frac{d}{d\tau}|_{\tau=0} F(u_{g(\tau)}X).
\l{x1}
\ee

It is shown in a standard way that:

1.  The quantity \r{x1} does not  depend  on  the
choice of the curve $g(\tau)$ with the tangent vector $A$.

2. The following property
$$
\delta[A_1+\alpha A_2] =  \delta[A_1]  +  \alpha  \delta[A_2],   \alpha
\in{\bf R}, A_1,A_2 \in T_e{\cal G}
$$
is satisfied.

Namely, let  $g_1(\tau)$ and $g_2(\tau)$ be smooth curves on the
Lie group ${\cal G}$ such that $g_1(0)=e$, $g_2(0)=e$. One has
$$
\frac{F(u_{g_1(\tau)g_2(\tau)} X) - F(X)}{\tau}   =   \int_0^1  d\xi
\frac{\partial}{\partial \tilde{\tau}}         F(u_{g_1(\tilde{\tau})}
u_{g_2({\tau})} X)|_{\tilde{\tau}         =         \xi\tau}         +
\frac{F(u_{g_2(\tau)}X)- F(X) }{\tau}.
$$
Considering the limit $\tau\to 0$, one obtains:
\be
\frac{d}{d\tau}|_{\tau=0} F(u_{g_1(\tau)g_2(\tau)} X) =
\frac{d}{d\tau}|_{\tau=0} F(u_{g_1(\tau)} X) +
\frac{d}{d\tau}|_{\tau=0} F(u_{g_2(\tau)} X).
\l{x1a}
\ee

Let $g(\tau)$ and $\tilde{g}(\tau)$ be curves on  $\cal  G$  with  the
tangent vector   $A$.  Choose  $g_1(\tau)  =  g(\tau)$,  $g_2(\tau)  =
g^{-1}(\tau) \tilde{g}(\tau)$.  Since $g_2(\tau)  -  e  =  O(\tau^2)$,
$\frac{d}{d\tau}|_{\tau=0} F(u_{g_2(\tau)}X)  =  0$.  It  follows from
eq\r{x1a} that
$\frac{d}{d\tau}|_{\tau=0} F(u_{g(\tau)}X)  =
\frac{d}{d\tau}|_{\tau=0} F(u_{\tilde{g}(\tau)}X)$.
Thus, definition \r{x1} is correct.

Furthermore,
let $g_1(\tau)$,  $g_2(\tau)$  be  curves with tangent vectors $A$ and
$B$ correspondingly.  Then $g_1(\tau)g_2(\tau)$ is a  curve  with  the
tangent vector $A+B$. Eq.\r{x1a} implies that
$$
\delta[A+B] = \delta[A] + \delta[B].
$$
Finally, consider  a  curve  $g(\tau)$ with the tangent vector $A$ and
the curve $\tilde{g}(\tau)= g(\alpha \tau)$ with  the  tangent  vector
$\alpha A$. One has
$$
\frac{d}{d\tau}|_{\tau=0} F(u_{\tilde{g}(\tau)}     X)     =    \alpha
\frac{d}{d\tau}|_{\tau=0} F(u_{g(\tau)}X).
$$
Thus, $\delta[\alpha A] =\alpha \delta[A]$.

Let $g\in{\cal G}$, $A\in T_e{\cal G}$, $h(\tau)$ be a curve on ${\cal
G}$  with tangent vector $B$ at $h(0)=e$.  Then the tangent vector for
the curve $gh(\tau)g^{-1}$ at $h(0)=e$ does not depend  on
the choice of the curve $h(\tau)$. Denote it by $gBg^{-1}$.

Define the operator $W_g$ on the space of functionals $F$ as
$W_gF[X] = F[u_{g^{-1}}X]$.

We see that the following property is satisfied:
\be
W_g \delta[B] W_{g^{-1}} F = \delta[gBg^{-1}]F.
\l{x3}
\ee

Let $g=g(\tau)$ be a curve with the tangent vector
$A$ at $g(0)=e$.  Differentiating expression
\r{x3} by $\tau$ at $\tau=0$, we obtain
the following relation:
\be
([\delta[A],\delta[B]] + \delta([A,B]))F = 0.
\l{x4}
\ee
Here $[A,B]$ is the Lie-algrbra commutator for the group $\cal G$.

Namely, let $g(\tau)$ be a smooth curve on the Lie group $\cal G$
with tangent vector $A$ at $g(0)=e$. Make use of the property \r{x3}:
\be
W_{g(\tau)} \delta[B]   W_{g^{-1}(\tau)}   =   \delta    [g(\tau)    B
g^{-1}(\tau)] F,
\l{x3a}
\ee
rewrite definition \r{x1} as
$$
\delta [A] = \frac{d}{d\tau}|_{\tau = 0} W_{g^{-1}(\tau)},
$$
remember that the Lie commutator can be defined as
$$
[A;B] \equiv \frac{d}{d\tau}|_{\tau=0} g(\tau) B g^{-1}(\tau).
$$
Consider the derivatives of sides of eq.\r{x3a}  at  $\tau  =  0$.  We
obtain property \r{x4}.

{\bf 2.}
Consider now the infinitesimal properties of the transformation
$U$. Suppose that on some dense subset $D$ of
$\cal F$ the vector functions  $U_g[X]\Psi \equiv U_g(u_gX\gets X)\Psi$
($\Psi\in D$)
are strongly continously differentiable  with respect to
$g$ and smooth with respect to $X$. Define operators
\be
H(A:X) \Psi = i\frac{d}{d\tau}|_{\tau=0} U_{g(\tau)}[X] \Psi,
\l{x4a}
\ee
where $g(\tau)$  is  a  curve  on  the group $\cal G$ with the tangent
vector $A$ at $g(0)=e$.

We find:\\
1. The operator  $H(A:X)$  does not depend on
the choice of the curve  $g(\tau)$   with the tangent vector $A$.\\
2. The following property is satisfied:
$$
H(A_1+\alpha A_2:X) = H(A_1:X) + \alpha H(A_2:X).
$$

Let $h(\tau)$ be a curve with the tangent vector $B$ at $h(0)=e$.
Eq.\r{x0} implies:
$$
U_g[u_{h(\tau)}X] U_{h(\tau)}[X]U_g^{-1}[X]\Psi  =
U_{gh(\tau)g^{-1}}[u_gX]\Psi, \qquad \Psi \in D
$$
Differentiating this identity by $\tau$ at $\tau=0$, we obtain that
for all $\Psi\in D$
\be
-iU_g[X] H(B:X) U_g^{-1}[X]\Psi + (\delta[B]U_g)[X] U_g^{-1}[X] \Psi =
-i H(gBg^{-1}: u_gX)\Psi.
\l{x5}
\ee

Let  $g=g(t)$  be a curve with the tangent vector
$A$ at $g(0)=e$. Differentiating eq.\r{x5} by $t$ in a weak sense,
we find that
on the subset $D$ the following bilinear form vanishes:
\be
-[H(A:X): H(B:X)]   -  i\delta[B] H(A:X) + i\delta[A] H(B:X)
+ i H([A;B]: X) = 0.
\l{x6}
\ee

{\bf 3.} Let us investigate now infinitesimal properties of  relations
\r{1.23} and \r{1.24}. They can be rewritten as
\be
\omega_{u_gX} [u_g^*(u_gX \gets X) \delta X] = \omega_X[\delta X];
\l{3.1}
\ee
\be
\Omega_{u_gX} [u_g^*(u_gX \gets X) \delta X] U_g(u_gX \gets X) =
U_g(u_gX\gets X) \Omega_X[\delta X].
\l{3.2}
\ee
It is  convenient  to  introduce  the  operator  $\delta[A]$  for  the
differential 1-forms.  Let  $\omega$   be   a   differential   1-form,
$g(\tau)$ be  a  curve  on  $\cal  G$  with  a  tangent  vector $A$ at
$g(0)=e$. Define:
$$
(\delta[A] \omega)_X [\delta X] \equiv \frac{d}{d\tau}|_{\tau=0}
\omega_{u_g(\tau) X} [u_{g(\tau)}^*(u_{g(\tau)}X \gets X) \delta X].
$$
Analogously,
$$
(\delta[A] \Omega)_X [\delta X] \equiv \frac{d}{d\tau}|_{\tau=0}
\Omega_{u_g(\tau) X} [u_{g(\tau)}^*(u_{g(\tau)}X \gets X) \delta X].
$$
Defining the operator $W_g$ as
$$
(W_{g^{-1}}\omega)_X [\delta X] =
\omega_{u_gX} [u_g^*(u_gX \gets X) \delta X] = \omega_X[\delta X],
$$
we check relations \r{x3}, \r{x4}.

Differentiating eqs.\r{3.1}, \r{3.2} for $g=g(\tau)$, we find:
\be
\delta[A] \omega = 0.
\l{3.3}
\ee
\be
(\delta[A] \Omega)_X[\delta X] - i [\Omega_X [\delta X]; H(A:X)] = 0.
\l{3.4}
\ee
Here $A$ is a tangent vector to $g(\tau)$ at $\tau = 0$.

\subsection{From algebras to groups}

Investigate now the problem of reconstructing the group representation
if the algebra representation is known.  Since classical invariance is
usually evident in applications,  we suppose that mappings $u_g: {\cal
X} \to  {\cal  X}$  which  satisfy  the condition \r{1.23} are already
specified. Our purpose is to reconstruct the unitary  operators  $U_g(
u_gX \gets  X)$  satisfying  the group property \r{x0},  provided that
operators $H(A:X)$ are known.

Without loss of generality, spaces ${\cal F}_X$ can be identified with
the space  $\cal  F$  because of local triviality of the semiclassical
bundle.

Let us impose the following conditions on the operators
$H(A:X)$,  $A  \in
T_e{\cal G}$, $X\in {\cal X}$.

G1. {\it Hermitian operators $H(A:X)$ are defined on a common
domain $\cal D$ which is dense in $\cal F$.
$\cal D$ is a subset of domains of the operators $\Omega_X[\delta X]$.
}

G2. {\it For each smooth curve $h(\alpha)$ on $\cal G$,  $\delta X \in
T_X{\cal X}$  and $\Psi
\in {\cal  D}$  the  vector  functions   $H(A:u_{h(\alpha)}X)\Psi$
and $\Omega_{u_{h(\alpha)}X} [u_{h(\alpha)}^*(u_{h(\alpha)}X
\gets X) \delta X]\Psi$ are
strongly continously differentiable with respect to
$\alpha$.}

G3. {\it The bilinear forms \r{x6} and \r{3.4} vanish on $\cal D$.}

Let ${Z} \in T_e {\cal G}$ be a subset of the Lie algebra of the
group $\cal G$ such that $span Z = T_e{\cal G}$. Let $B \in {Z}$,  while
$g_B(t)$  is a one-parametric subgroup of the Lie group
$\cal  G$ with the tangent vector
$B$  at  $t=0$,  $g_B(0)=e$.
Denote by $U_B^t(X)$ the operator taking the initial condition
$\Psi_0 \in {\cal D}$ of the Cauchy problem for the equation
\be
i \frac{\partial\Psi_t}{\partial t} = H(B:u_{g_B(t)}X) \Psi_t
\l{x8}
\ee
($\partial\Psi_t/\partial t$  is a strong derivative)
to the solution  $\Psi_t
\in {\cal  D}$ of the Cauchy problem,
$\Psi_t =  U_B^t(X)\Psi_0$.  This  definition  is  correct  under  the
following condition.

G4. {\it  Let $B \in {Z}$.
If  $\Psi_0\in  {\cal D}$, there exists a solution
of the Cauchy problem for eq.\r{x8}. }

Uniqueness of the solution is a corollary of the property
$||\Psi_t||  =
||\Psi_0||$ which is checked directly by differentiation.
The isometric operator  $U_B^t(X)$ can be  extended
then from  ${\cal D}$ to $\cal F$. It satisfies the
property
$$
U_B^{t_1}(u_{g_B(t_2)}X) U_B^{t_2}(X) = U_B^{t_1+t_2} (X).
$$
Therefore, it is invertible and unitary.

Impose also the following conditions.

G5. {\it Let $B\in {Z}$.
For each smooth curve $h(\alpha)$ on  $\cal  G$  and each
$\Psi_0\in {\cal   D}$ the quantity $||\frac{\partial}{\partial\alpha}
H(A:u_{h(\alpha)}X) \Psi_t||$ is bounded uniformly
with respect to  $\alpha,t  \in
[\alpha_1,\alpha_2] \times [t_1,t_2]$ for any finite
$\alpha_1,\alpha_2,t_1,t_2$.
}

G6. {\it For $\Psi \in {\cal D}$, $B\in {Z}$, $A\in T_e{\cal G}$,
$\delta X \in T_X{\cal X}$,
the following properties are satisfied:
\be
|| H(g_B(\tau)Ag_B^{-1}(\tau): u_{g_B(\tau)} X)
[U_B^{\tau}(X) \Psi - \Psi]
|| \to_{\tau\to 0} 0;
\l{3.5a}
\ee
\be
||\Omega_{u_{g_B(\tau)} X} [u_{g_B(\tau)}^*(u_{g_B(\tau)}X \gets X)
\delta X] [U_B^{\tau}(X) \Psi - \Psi]
|| \to_{\tau\to 0} 0.
\l{3.5}
\ee
}

{\bf Lemma 3.1.} {\it Let $B \in Z$, $\delta X \in T_X {\cal X}$. Then
the property
\be
A^t \equiv
(U^t_B(X))^{-1}
\Omega_{u_{g_B(t)} X} [u_{g_B(t)}^*(u_{g_B(t)}X \gets X) \delta X]
U^t_B(X) - \Omega_X[\delta X] = 0
\l{3.6}
\ee
is satisfied on the domain $\cal D$.
}

{\bf Proof.}  Let us consider the matrix element of the left-hand side
of eq.\r{3.6}.  At $t=0$  it  vanishes.  Let  us  calculate  its  time
derivative,
\beb
\frac{1}{\tau} [(\Phi, A^{t+\tau} \Psi)  - (\Phi, A^t \Psi)] =\\
\left(\Omega_{u_{g_B(\tau)}X_t}
[u_{g_B(\tau)}^*(u_{g_B(\tau)}X_t \gets  X_t) \delta X_t]
\Phi_{t+ \tau}, \frac{\Psi_{t+\tau} - \Psi_t}{\tau}\right) + \\
\left(\Phi_{t+\tau},
\frac{
\Omega_{u_{g_B(\tau)}X_t}
[u_{g_B(\tau)}^*(u_{g_B(\tau)}X_t \gets  X_t) \delta X_t]
- \Omega_{X_t}[\delta X_t]
}{\tau} \Psi_t\right) + \\
\left(\frac{\Phi_{t+\tau}  - \Phi_t}{\tau},
\Omega_{X_t} [\delta X_t] \Psi_t \right),
\eeb
where $\Phi_t \equiv U_B^t(X) \Phi$,
$\Psi_t \equiv U_B^t(X) \Psi$, $X_t = u_{g_B(t)}X$,
$\delta X_t = u_{g_B(t)}^*(u_{g_B(t)}X \gets X) \delta X$.
It follows  form  eq.\r{3.4} and property G6 that $\frac{d}{dt} (\Phi,
A^t \Psi) = 0$,  so that expression \r{3.6}  vanishes  as  a  bilinear
form. Since  it  is  defined  on $\cal D$,  it vanishes as an operator
expression. Lemma 3.1 is proved.

Under conditions G1-G6, we obtain:

{\bf Lemma 3.2.} {\it Let $B\in {Z}$, $A\in T_e{\cal G}$,
The following property is satisfied on the domain
$\cal D$:
\be
H(A:X) +   i  (U^t_B(X))^{-1}  (\delta[A]U_B^t)(X)  -  (U^t_B(X))^{-1}
H(g_B(t)A g_B(t)^{-1}: u_{g_B(t)} X) U_B^t(X) = 0.
\l{x9}
\ee
}

{\bf Proof.} Let us check that under  these  conditions  the  operator
$(\delta[A]U_B^t)(X)$ is correctly defined, i.e. the strong derivative
\be
(\delta[A] U_B^t)(X)  \Psi  =  \frac{d}{d\alpha}|_{\alpha  =  0} U_B^t
(u_{h(\alpha)} X) \Psi
\l{x9+}
\ee
exists for  all  $\Psi \in {\cal D}$,  where $h(\alpha)$ is a curve on
$\cal G$ with tangent vector $A$.

Denote
$$
\Psi_{\alpha, t} = U_B^t(u_{h(\alpha)}X) \Psi.
$$
This vector obeys the equation
$$
i \frac{\partial}{\partial     t}     \Psi_{\alpha,t}      =      H(B:
u_{g_B(t)h(\alpha)} X) \Psi_{\alpha,t},
$$
so that
\beb
i \frac{\partial}{\partial  t}
(\Psi_{\alpha+  \delta  \alpha,  t } - \Psi_{\alpha,t})
= H(B:u_{g_B(t) h(\alpha+\delta\alpha)} X)
(\Psi_{\alpha+  \delta  \alpha,  t } - \Psi_{\alpha,t})
\crr
+ (H(B:  u_{g_B(t)  h(\alpha+\delta  \alpha)}X)  - H(B:  u_{g_B(t)}X))
\Psi_{\alpha,t}.
\eeb
Since $\Psi_{\alpha,0} = \Psi_{\alpha+\delta\alpha,0} = \Psi$, we have
$$
\Psi_{\alpha+\delta \alpha,  t} - \Psi_{\alpha,t} = - i \int_0^t d\tau
U_B^{t-\tau} (u_{g_B(\tau) h(\alpha+\delta \alpha)} X)
(H(B: u_{g_B(\tau) h(\alpha+\delta \alpha)}X) - H(B:  u_{g_B(\tau)}X))
\Psi_{\alpha,\tau}.
$$
Because of   unitarity   of   the  operators  $U_B^t$,  the  following
estimation takes place:
$$
||\Psi_{\alpha+\delta \alpha,  t} - \Psi_{\alpha,t}||  \le
\int_0^t d\tau
|| (H(B: u_{g_B(\tau) h(\alpha+\delta \alpha)}X) - H(B:  u_{g_B(\tau)}X))
\Psi_{\alpha,\tau}||.
$$
Making use  of  the Lesbegue theorem \c{KF} and condition G5,  we find
that $||\Psi_{\alpha+\delta \alpha,t} - \Psi_{\alpha,t}||  \to_{\delta
\alpha =  0}  0$,  so  that  the  operator  $U_B^t(u_{h(\alpha)}X)$ is
strongly continous with respect to $\alpha$.

Furthermore,
$$
\frac{\Psi_{\alpha +  \delta   \alpha,t}   -   \Psi_{\alpha,t}}{\delta
\alpha} =  - i\int_0^t d{\tau} \int_0^1 d\gamma U_B^{t-\tau}(u_{g_B(\tau)
h(\alpha+ \delta \alpha)} X ) \left(
\frac{\partial}{\partial \alpha}  H(B:  u_{g_B(\tau) h(\alpha + \gamma
\delta \alpha)}X)
\right) \Psi_{\alpha, \tau}
$$
Denote
\be
\frac{\partial \Psi_{\alpha,t}}{\partial  \alpha}  \equiv - i \int_0^t
d\tau U_B^{t-\tau}(u_{g_B(\tau)
h(\alpha+ \delta \alpha)} X ) \left(
\frac{\partial}{\partial \alpha}  H(B:  u_{g_B(\tau) h(\alpha)}X)
\right) \Psi_{\alpha, \tau}.
\l{x9b}
\ee
The following estimation takes place:
\bea
|| \frac{\Psi_{\alpha  + \delta \alpha,  t} - \Psi_{\alpha,t} }{\delta
\alpha} -  \frac{\partial  \Psi_{\alpha,t}}{\partial  \alpha}  ||  \le
\int_0^t d\tau \int_0^1 d\gamma
\left(
||U_B^{t-\tau} (u_{g_B(\tau) h(\alpha+ \delta \alpha)}X)||
\right. \crr \times
||\left[
\frac{\partial H}{\partial   \alpha}  (B:  u_{g_B(\tau)h(\alpha+\gamma
\delta \alpha)}X) -
\frac{\partial H}{\partial   \alpha}  (B:  u_{g_B(\tau)h(\alpha)}X)
\right]
\Psi_{\alpha,\tau}||
+
 \crr \left.
||(U_B^{t-\tau} (u_{g_B(\tau) h(\alpha+ \delta \alpha)}X)
- U_B^{t-\tau} (u_{g_B(\tau) h(\alpha)}X))
\frac{\partial H}{\partial   \alpha}  (B:  u_{g_B(\tau)h(\alpha)}X)
\Psi_{\alpha,\tau}||
\right)
\l{x9a}
\eea
Making use of the Lesbegue theorem, conditions G2,G5, we find that the
quantity \r{x9a} tends to zero as $\delta \alpha  \to  0$.  Thus,  the
vector \r{x9+} is correctly defined.

It follows from the expression \r{x9b} that
$$
\frac{\partial}{\partial t}    \frac{\partial}{\partial     \alpha}
U_B^t(u_{h(\alpha)} X) =
-i \frac{\partial}{\partial     \alpha}
H(B: u_{g_B(t) h(\alpha)}X) \cdot
U_B^t(u_{h(\alpha)} X) - i H(B: u_{g_B(t) h(\alpha)}X)
\frac{\partial U_B^t (u_{h(\alpha)}X)}{\partial \alpha}
$$
in a strong sense.

Let us prove now property \r{x9}.
At $t=0$ the property \r{x9} is satisfied. The derivative with
respect to $t$ of any matrix element of the
operator \r{x9} under conditions H1-H6 vanishes.
Therefore, equality  \r{x9}  viewed  in  terms  of  bilinear  forms is
satisfied on $\cal D$. Since the left-hand side of
eq.\r{x9} is defined on $\cal D$, it also vanishes on $\cal D$.

Let the property
\be
g_{B_n}(t_n(\alpha)) ... g_{B_1} (t_1(\alpha)) = e,
\l{x11}
\ee
be satisfy for $\alpha \in [0,\alpha_0]$ and
$B_1, ..., B_n\in {Z}$. Here
$t_k(\alpha)$ are smooth functions. Denote $h_k(\alpha) =
g_{B_k}(t_k(\alpha))$, $s_k(\alpha) = h_k(\alpha) ... h_1(\alpha)$.

{\bf Lemma 3.3.} {\it Under condition \r{x11} the operator
\be
U_{B_n}^{t_n(\alpha)}(u_{s_{n-1}(\alpha)}X)
...
U_{B_1}^{t_1(\alpha)}(X)
\l{x12}
\ee
is $\alpha$-independent.
}

To prove lemma, denote
$U_k \equiv U_k(u_{s_{k-1}(\alpha)}X) \equiv
U_{B_k}^{t_k(\alpha)}(u_{s_{k-1}(\alpha)}X)$.
Let us use the following lemma.

{\bf Lemma 3.4.} {\it Let $s(\alpha)$ be a smooth curve on the group
$\cal G$,  $t(\alpha)$ is a smooth real function,  $B\in
T_e{\cal G}$. Then the operator function
$U_B^{t(\alpha)} (u_{s(\alpha)}X)$    is    strongly    differentiable
with respect to $\alpha$ on $\cal D$ and
\be
\frac{\partial}{\partial\alpha} U_B^{t(\alpha)} (u_{s(\alpha)}X) = - i
\frac{dt}{d\alpha} H(B:u_{g_B(t(\alpha)) s(\alpha)}X)
U_B^{t(\alpha)}
(u_{s(\alpha)}X) + (\delta[\frac{ds}{d\alpha} s^{-1}]
U_B^{t(\alpha)}) (u_{s(\alpha)}X),
\l{x10}
\ee
where
$\frac{ds}{d\alpha} s^{-1}$ is a
tangent vector to the curve
$s(\alpha+\tau)s^{-1}(\alpha)$ at $\tau=0$.
}

{\bf Proof.}
Let $\Psi \in {\cal D}$. One has
\bea
\frac{1}{\delta \alpha}
(U_B^{t(\alpha+\delta
\alpha)}(u_{s(\alpha+\delta \alpha)}X)
- U_B^{t(\alpha)}(u_{s(\alpha)}X)) \Psi = \crr
\frac{1}{\delta \alpha}
(U_B^{t(\alpha+\delta
\alpha)}(u_{s(\alpha)}X)
- U_B^{t(\alpha)}(u_{s(\alpha)}X)) \Psi
+
\frac{1}{\delta \alpha}
(U_B^{t(\alpha+\delta
\alpha)}(u_{s(\alpha+\delta \alpha)}X)
- U_B^{t(\alpha+ \delta \alpha)}(u_{s(\alpha)}X)) \Psi.
\l{x10a}
\eea
The first term in the right-hand side of eq.\r{x10a} tends to
$$
-i \frac{dt}{d\alpha} H(B:u_{g_B(t(\alpha)) s(\alpha)}X) \Psi
$$
by definition of the operator $U_B^t(X)$. Consider the second term. It
can be represented as
$$
\int_0^1 d\gamma           \frac{\partial}{\partial            \alpha}
U_B^{t(\overline{\alpha})} (u_{s(\alpha      +      \gamma      \delta
\alpha)}X)|_{\overline{\alpha} = \alpha + \delta \alpha} \Psi.
$$
Making use of eq. \r{x9b}, we take this term to the form
\be
-i \int_0^1 d\gamma \int_0^{t(\alpha+\delta \alpha)} d\tau
U_B^{t(\alpha+\delta\alpha)-\tau} (u_{g_B(\tau)    s(\alpha+    \gamma
\delta \alpha)}           X)           \frac{\partial}{\partial\alpha}
H(B:u_{g_B(\tau)s(\alpha+ \gamma\delta \alpha}) X)  U_B^{\tau}
(u_{s(\alpha + \gamma \delta \alpha)}X) \Psi.
\l{x10b}
\ee
Making use of the Lesbegue theorem and property G6,  we see  that  the
vector \r{x10b} is strongly continous as $\delta\alpha \to 0$, so that
it is equal to
$$
\frac{\partial}{\partial \alpha}            U_B^{t(\overline{\alpha})}
(u_{s(\alpha)}X)|_{\overline{\alpha}= \alpha} \Psi =
((\delta[\frac{ds}{d\alpha}s^{-1}] U_B^{t(\alpha)})  (u_{s(\alpha)} X)
\Psi.
$$
We obtain formula \r{x10}.

Let us return to proof of lemma 3.3.
To check formula \r{x12}, let us obtain that
\be
\frac{d}{d\alpha} (U_n...U_1) (U_n...U_1)^{-1} = 0
\l{x13}
\ee
on $\cal  D$  (the  derivative  is  viewed  in the strong sense).  The
property \r{x13} is equivalent to the folowing relation:
\be
\sum_{k=1}^n U_n...U_{k+1}    \frac{\partial   U_k}{\partial   \alpha}
U_k^{-1}...U_n^{-1} = 0.
\l{x14}
\ee
Making use of eq.\r{x10}, we take eq.\r{x14} to the form
\bea
\sum_{k=1}^n U_n...U_{k+1}  H(-i\frac{dt_k}{d\alpha}  B_k :  u_{s_k}X)
U_{k+1}^{-1} ...   U_n^{-1}   +  \crr \sum_{k=1}^n   U_n    ...    U_{k+1}
(\delta[\frac{ds_{k-1}}{d\alpha}       s_{k-1}^{-1}]U_k)(u_{s_{k-1}}X)
U_k^{-1} ... U_n^{-1}=0.
\l{x15}
\eea
Applying properties \r{x9} $n-k$ times, we obtain
\bea
-iH(\sum_{k=1}^n \frac{dt_k}{d\alpha} h_n ... h_{k+1} B_k h_{k+1}^{-1}
... h_n^{-1}: X) + \sum_{l=1}^n U_n ... U_{l+1}
(\delta[\frac{ds_{l-1}}{d\alpha}       s_{l-1}^{-1}]U_l)
(u_{s_{l-1}}X)
U_l^{-1} ... U_n^{-1}
\crr
- \sum_{k=1}^n  \sum_{l=k+1}^n  U_n  ...  U_{l+1} \frac{dt_k}{d\alpha}
\delta[h_{l-1}... h_{k+1} B_k h_{k+1}^{-1} ... h_{l-1}^{-1}]U_l)
(u_{s_{l-1}}X)
U_l^{-1} ... U_n^{-1}.
\l{x16}
\eea
Eq.\r{x11} implies that
\beb
\sum_{k=1}^n \frac{dt_k}{d\alpha} h_n ... h_{k+1} B_k h_{k+1}^{-1}
... h_n^{-1} = 0; \crr
\frac{ds_{l-1}}{d\alpha} s_{l-1}^{-1} - \sum_{k=1}^{l-1}
h_{l-1}... h_{k+1}     B_k     h_{k+1}^{-1}      ...      h_{l-1}^{-1}
\frac{dt_k}{d\alpha}
= 0.
\eeb
Lemma 3.3 is proved.

{\bf Corollary.} {\it Let $t_k(0)=e$. Then
$$
U_{B_n}^{t_n(\alpha)}(u_{s_{n-1}(\alpha)}X)
...
U_{B_1}^{t_1(\alpha)}(X) = 1
$$
under conditions of lemma 3.3.
}

{\bf Theorem 3.5.} {\it Under conditions G1-G6 semiclassical bundle is
invariant under local Lie group $\cal G$.}

{\bf Proof.} Let $B_1,...,B_m$ be a basis on the Lie algebra. Then one
can uniquely  introduce  the  canonical coordinates of the second kind
\c{Pontriagin} on the local Lie group by the formula
$$
g = g_{B_1}(\alpha_1) ... g_{B_m} (\alpha_m).
$$
Set
$$
U_g(u_gX\gets X) =
U_{B_1}^{\alpha_1}
(u_{g_{B_2}(\alpha_2) ... g_{B_m} (\alpha_m)}X)
U_{B_2}^{\alpha_2}
(u_{g_{B_3}(\alpha_3) ... g_{B_m} (\alpha_m)}X)
...
U_{B_m}^{\alpha_m}(X).
$$
The group property is then a corollary of lemma 3.3. Namely, let
\beb
g(t) = g_{B_1}(t\alpha_1) ... g_{B_m} (t\alpha_m);\\
h(t) = g_{B_1}(t\beta_1) ... g_{B_m} (t\beta_m).
\eeb
Then the relation
$$
g(t)h(t) = g_{B_1}(\gamma_1(t)) ... g_{B_m} (\gamma_m(t))
$$
specifies the   coordinates  $\gamma_i(\alpha,\beta,t)$  in  a  unique
fashion, since the second-kind  canonical  coordinates  are  correctly
defined \c{Pontriagin}. Moreover, the dependence
$\gamma_i(\alpha,\beta,t)$ is smooth.

It follows then from lemma 3.3 that
$$
(U_{g(t)h(t)} [u_{g(t)h(t)} X \gets X])^{-1}
U_{g(t)} [u_{g(t)h(t)} X \gets u_{h(t)} X]
U_{h(t)} [u_{h(t)} X \gets X] = 1.
$$
Theorem is proved.

\section{Quadratic infinitesimal operators in the
Fock space}

Symmetries of quantum mechanical systems can be investigated "exactly"
without any  approximations.  Realistic  models   of   QFT   are   not
constructed mathematically,  so  that  investigations of semiclassical
field theory may give rise  to  surprising  results  such  as  quantum
anomalies. In this section conditions G1-G6 are reformulated, provided
that ${\cal F}_X$ are Fock spaces,  $H(A:X)$ are quadratic in creation
and annihilation operators in the Fock space, $\Omega_X[\delta X]$ are
linear combinations of creation and annihilation operators.

Remind that the Fock space ${\cal F}(L^2({\bf R}^l))$ is defined  as  a
space of sets
$$
\Psi= (\Psi_0,  \Psi_1({\bf  x}_1),  ...  ,\Psi_n({\bf   x}_1,...,{\bf
x}_n),...)
$$
of symmetric  with  respect to $x_1$,  ...,  $x_n$ symmetric functions
$\Psi_n$ such that $||\Psi||^2 =  \sum_{n=0}^{\infty}  |\Psi_n||^2  <
\infty$. By $A^{\pm}({\bf x})$ we denote,  as usual,  the creation and
annihilation operator distributions:
\beb
(\int d{\bf  x}  A^+({\bf  x})  f({\bf x}) \Psi)_n ({\bf x}_1,...,{\bf
x}_n) = \frac{1}{\sqrt{n}} \sum_{l=1}^n f({\bf x}_j) \Psi_{n-1}  ({\bf
x}_1,..., {\bf x}_{j-1}, {\bf x}_{j+1}, ..., {\bf x}_n); \crr
(\int d{\bf x} A^-({\bf x})  f^*({\bf x}) \Psi)_{n-1}
({\bf x}_1,...,{\bf  x}_{n-1})  =  \sqrt{n} \int d{\bf x} f^*({\bf x})
\Psi_n({\bf x},{\bf x}_1,...,{\bf  x}_{n-1}).
\eeb
By $|0>$   we  denote,  as  usual,  the  vacuum  vector  of  the  form
$(1,0,0,...)$. Arbitrary vector of the Fock space can be presented via
the creation operators and vacuum vector as follows
\c{Ber}
$$
\Psi = \sum_{n=0}^{\infty} \frac{1}{\sqrt{n!}}
\int d{\bf x}_1 ...  d{\bf x}_n \Psi_n({\bf
x}_1,...,{\bf x}_n) A^+({\bf x}_1) ... A^+({\bf x}_n)|0>.
$$
Introduce the  operator  of  number of particles $\hat{n}$ as $(\hat{n}
\Psi)_n = n \Psi_n$.

Let ${\cal  H}^{\pm\pm}$ be operators in $L^2({\bf R}^l)$ with kernels
${\cal H}^{\pm\pm}({\bf    x},{\bf    y})$.    By
$A^{\pm}    {\cal H}^{\pm\pm} A^{\pm}$
we denote the operators in the Fock space
$$
A^{\pm}    {\cal H}^{\pm\pm} A^{\pm}
= \int d{\bf x} d{\bf y}
A^{\pm}({\bf x})  {\cal H}^{\pm\pm}({\bf x},{\bf y})A^{\pm}({\bf y}).
$$
For the case of an unbounded operator ${\cal  H}^{+-}$,  the  operator
$A^+ {\cal H}^{+-} A^-$ can be defined as
$$
(A^+{\cal H}^{+-} A^- \Psi)_n =
\sum_{j=1}^n  1^{\otimes  j-1}  \otimes {\cal H}^{+-}  \otimes
1^{\otimes n-j} \Psi_n.
$$
Denote also
$$
A^{\pm} \varphi = \int d{\bf x} A^{\pm}({\bf x}) \varphi({\bf x}).
$$
Let
\be
H(B:X) =
\frac{1}{2} A^+ {\cal H}^{++} (B:X) A^+
+   A^+ {\cal H}^{+-} (B:X) A^-
+ \frac{1}{2} A^- {\cal H}^{--} (B:X) A^-
+ \overline{\cal H}(B:X);
\l{4.1a}
\ee
where $({\cal H}^{++})^+ = {\cal H}^{--}$, $({\cal H}^{+-})^+ =
{\cal H}^{+-}$;
\be
\Omega_X[\delta X]     =     -    i    (A^+\varphi_X[\delta    X]    -
A^-\varphi_X^*[\delta X])
\l{4.1}
\ee
for some $L^2({\bf R}^l)$-valued 1-form $\varphi$.

Let us  formulate  the  conditions  that are sufficient for satisfying
properties G1-G6.

In QFT-applications \c{Shvedov},  the operators  ${\cal  H}^{+-}(B:X)$
are unbounded. However, the singular unbounded part is $X$-independent,
$$
{\cal H}^{+-}(B:X) = L(B) + {\cal H}(B:X),
$$
while ${\cal H}(B:X)$ is a bounded operator.

Impose the following conditions on the self-adjoint operator $L$.

{\bf F1.} {\it
There exists a positively  definite  self-adjoint  operator  $T$  such
that:\\
(i) $||T^{-1/2} L(B) T^{-1/2}|| <\infty$,
$||L(B) T^{-1}|| < \infty$;\\
(ii) for all $t_0$ there exists such a constant $C$ that
$||T^{1/2} e^{-iL(B)t} T^{-1/2} || \le C$,
$||T e^{-iL(B)t} T^{-1} || \le C$ for $t\in (-t_0,t_0)$.
}

{\bf F2}. {\it  For any smooth curve $h(\alpha)$ on $\cal G$:\\
(i) the function $\overline{\cal H}(B: u_{h(\alpha)}X)$ is continously
differentiable;\\
(ii) the operator-valued function
${\cal H}^{++}(B: u_{h(\alpha)}X)$
is continously differentiable in the Hilbert-Schmidt norm
$||O||_2 = \sqrt{Tr O^+O}$;\\
(iii) the operator functions
$T^{-1/2} {\cal H}^{+-}(B: u_{h(\alpha)}X) T^{-1/2}$
and
${\cal H}^{+-}(B:     u_{h(\alpha)}X)    T^{-1}$    are    continously
differentiable in the operator norm $||\cdot||$;\\
(iv) the operator function
$T{\cal H}^{++}(B:    u_{h(\alpha)}X)$    is    continous    in    the
Hilbert-Schmidt norm;\\
(v) the operator functions
$T {\cal H}(B: u_{h(\alpha)}X) T^{-1}$,
$T^{1/2} {\cal H}(B: u_{h(\alpha)}X) T^{-1/2}$ and
${\cal H}(B: u_{h(\alpha)}X)$ are strongly continous;\\
(vi) the function
$\varphi_{u_{h(\alpha)}X} [u_{h(\alpha)}^*   [u_{h(\alpha)}X   \gets  X)
\delta X] \in L^2({\bf R}^l)$ is strongly continously differentiable.
}

{\bf F3.} {\it The following commutation relations are satisfied:
\beb
{\cal H}^{++}([A;B]:X) = - i
[{\cal H}^{+-}(A:X) {\cal H}^{++}(B:X)
+ {\cal H}^{++}(B:X) ({\cal H}^{+-}(A:X))^* \\
- {\cal H}^{+-}(B:X) {\cal H}^{++}(A:X)
- {\cal H}^{++}(A:X) ({\cal H}^{+-}(B:X))^*]
+ \delta[A] {\cal H}^{++}(B:X)
- \delta[B] {\cal H}^{++}(A:X);\\
{\cal H}^{+-}([A;B]:X) = - i
[{\cal H}^{++}(B:X) ({\cal H}^{++}(A:X))^*
- {\cal H}^{++}(A:X) ({\cal H}^{++}(B:X))^* \\
+ [ {\cal H}^{+-}(A:X);  {\cal H}^{+-}(B:X)]]
+ \delta[A] {\cal H}^{+-}(B:X)
- \delta[B] {\cal H}^{+-}(A:X);\\
\overline{\cal H}([A;B]:X) =
- \frac{i}{2} Tr
[{\cal H}^{++}(B:X) ({\cal H}^{++}(A:X))^*
- {\cal H}^{++}(A:X) ({\cal H}^{++}(B:X))^*] \\
+ \delta[A] \overline{\cal H}(B:X)
- \delta[B] \overline{\cal H}(A:X)
\eeb
in a sense of bilinear forms in $D(T)$;
\beb
i (\delta[A]\varphi)_X[\delta X] = {\cal H}^{+-}(A:X) \varphi_X[\delta
X] + {\cal H}^{++}(A:X) \varphi^*_X[\delta X].
\eeb
}

Note that  condition  F1  is  an  alternative  for known conditions of
integrability of Lie algebra representations \c{BR,N,FS1,FS2,FS3}.

Let us check properties G1-G6.

\subsection{Some properties of the Fock space}

Without loss of generality,  one can suppose that $T-1>0$.  Otherwise,
one can change $T \to T+1$; properties F1-F3 will remain valid then.

Introduce the following norms in the Fock space:
$$
||\Psi||_m = ||(\hat{n}+1)^m \Psi||, \qquad
||\Psi||^T_m = || (A^+TA^-+1)^m \Psi||.
$$

{\bf Lemma 4.1.} {\it Let $||\Psi||^T_m <\infty$.  Then $||\Psi||_m \le
||\Psi||_m^T$.}

{\bf Proof.} It is sufficient to check that
$$
(\Psi_n, (\sum_{j=1}^n 1^{\otimes j-1} \otimes  T  \otimes  1^{\otimes
n-j} +1)^{2m} \Psi_n ) \ge (\Psi_n, (\sum_{j=1}^n 1 + 1)^{2m} \Psi_n).
$$
This relation is a corollary of the formula
$$
(\Psi_n, T^{2l_1}   \otimes   ...   \otimes   T^{2l_n}   \Psi_n)   \ge
(\Psi_n,\Psi_n)
$$
for all $l_1,...,l_n\ge 0$.  The latter formula is obtained  from  the
relation $||T^{-l_1} \otimes ...  \otimes T^{-l_n} || \le 1$. Lemma B.1
is proved.

Let ${\cal H}^{+-}$ be a nonbounded operator in $L^2({\bf R}^l)$  such
that operators $T^{-1/2}{\cal H}^{+-} T^{-1/2}$ and
${\cal H}^{+-}T^{-1}$ are bounded.

{\bf Lemma 4.2.} {\it The following estimation is satisfied:
$$
||A^+{\cal H}^{+-}A^- \Psi|| \le C ||\Psi||^T_1
$$
with $C=max(||T^{-1/2}{\cal         H}^{+-}         T^{-1/2}||,||{\cal
H}^{+-}T^{-1}||)$.
}

{\bf Proof.} One should check
\be
(\Psi_n,{\cal H}^{+-}_i{\cal H}^{+-}_j \Psi_n) \le C
(\Psi_n, T_iT_j \Psi_n)
\l{bb1}
\ee
with ${\cal H}^{+-}_i = 1^{i-1} \otimes {\cal H}^{+-} \otimes 1^{n-i}$,
$T_i =     1^{i-1}     \otimes     T    \otimes    1^{n-i}$.    Denote
$T_i^{1/2}T_j^{1/2}\Psi_n =\Phi_n$. Inequality \r{bb1} takes the form
\be
(\Phi_n, T_i^{-1/2}  T_j^{-1/2}   {\cal   H}^{+-}_i   {\cal   H}^{+-}_j
T_i^{-1/2} T_j^{-1/2} \Phi_n) \le C^2 (\Phi_n,\Phi_n).
\l{bb2}
\ee
For $i\ne j$, property \r{bb2} is satisfied if $C = ||T^{-1/2}{\cal
H}^{+-} T^{-1/2}||$  as a corollary of the Cauchy-Bunyakovski-Schwartz
inequality. For $i=j$,  property \r{bb2} is  satisfied  if  $C=||{\cal
H}^{\pm}T^{-1}||$. Lemma 4.2 is proved.

{\bf Lemma 4.3.} {\it Consider the operator
$$
\hat{\varphi} =  \int d{\bf x}_1 ...  d{\bf x}_m d{\bf y}_1 ...  d{\bf
y}_k \varphi  ({\bf  x}_1,...,{\bf  x}_n,  {\bf  y}_1,...,{\bf   y}_k)
A^+({\bf x}_1) ... A^+({\bf x}_m) A^-({\bf y}_1) ... A^-({\bf y}_k)
$$
with $\varphi \in L^2({\bf R}^{l(m+k)})$.  Let $\Psi \in {\cal F}$ and
$||\Psi||_{l+\frac{k+m}{2}} < \infty$. Then
$$
||\hat{\varphi} \Psi||_l  \le  C   ||\varphi||_{L^2}   ||\Psi||_{l   +
\frac{k+m}{2}},
$$
where $C^2 = \max\{1,(m-k)!(m-k)^{2l}\}$.
}

{\bf Proof.} One has
\beb
(\hat{\varphi}\Psi)_n({\bf z}_1,...,{\bf z}_n) =
\sqrt{\frac{(n-m+k)!}{(n-m)!}} \sqrt{\frac{n!}{(n-m)!}}
Sym \int d{\bf y}_1 ... d{\bf y}_k \varphi({\bf z}_1,...,{\bf z}_m,
{\bf y}_1,...,{\bf  y}_k) \crr \times
\Psi_{n-m+k}({\bf y}_1,...,{\bf y}_k,  {\bf
z}_{m+1},...,{\bf z}_n)
\eeb
where $Sym$ is a symmetrization operator.  Since $||Sym  \Phi_n||  \le
||\Phi_n||$ and
$$
||\int dy   \varphi({\bf   z},{\bf  y})  \Psi({\bf  y},{\bf  z}')||  \le
||\varphi|| ||\Psi||,
$$
one has
$$
||(\hat{\varphi}\Psi)_n|| \le
\sqrt{\frac{(n-m+k)!}{(n-m)!}} \sqrt{\frac{n!}{(n-m)!}}
||\varphi||
||\Psi_{n-m+k}||.
$$
Therefore,
\beb
||\hat{\varphi}\Psi||_l^2 =       \sum_{n=0}^{\infty}       (n+1)^{2l}
\frac{(n-m+k)!}{(n-m)!} \frac{n!}{(n-m)!}                ||\varphi||^2
||\Psi_{n-m+k}||^2 \crr
= \sum_{s=0}^{\infty}  \frac{(s+m-k)^{2l}  (s-k+1)  ...  s (s-k+1) ...
(s-k+m)}{(s+1)^{2l+k+m}} ||\varphi|| ||\Psi_s||^2 (s+1)^{2l+k+m} \crr
\le C^2 |\varphi||^2 ||\Psi||^2_{l+\frac{k+m}{2}},
\eeb
where $s=m-n+k$. Lemma is proved.

{\bf Corollary.}
\beb
||A^{\pm} \varphi    \Psi||   \le   ||\varphi||   ||\Psi||_{1/2}   \le
||\varphi|| ||\Psi||_1;\\
||\frac{1}{2} A^{\pm}    {\cal    H}^{\pm    \pm}    A^{\pm} \Psi||
\le
\frac{1}{\sqrt{2}} ||{\cal H}^{\pm \pm}||_2 ||\Psi||_1.
\eeb

Therefore, we obtain the following result.

{\bf Lemma 4.4.} {\it Let properties F1,  F2 be satisfied.  Set ${\cal D} =
\{\Psi \in  {\cal  F} | ||\Psi||_1^T \le \infty \}$.  Let the solution
for the Cauchy problem for eq.\r{x8} exists for all $\Psi_0 \in  {\cal
D}$ and  $\Psi_t$  be  continous  in  the $||\cdot||_1^T$-norm.  Then
properties G1,G2, G5, G6 are satisfied.
}

The proof is straightforward.

\subsection{Evolution with quadratic Hamiltonians}

Since property  G3 is a direct corollary of H3,  the remaining part of
checking conditions G1-G6 is to prove that the Cauchy problem for  the
equation
\bea
i \frac{d\Psi_t}{dt} = H_t\Psi_t,
\crr
H_t = H(B:u_{g_B(t)}X) =
\frac{1}{2} A^+{\cal  H}^{++}_t  A^-  +  A^+  {\cal  H}^{+-}_t  A^-  +
\frac{1}{2} A^- {\cal H}^{--}_t A^- + \overline{\cal H}_t.
\l{bb3}
\eea
on the  Fock  vector  $\Psi_t$ is correct and $\Psi_t$ is continous in
$||\cdot||_1^T$-norm. The strong derivative enters to eq.\r{bb3}.

Formally, the solution for the initial condition
\be
\Psi_0 = \sum_{n=0}^{\infty} \frac{1}{\sqrt{n!}} \int d{\bf  x}_1  ...
d{\bf x}_n   A^+({\bf   x}_1)  ...  A^+({\bf  x}_n)  \Psi_{0,n}  ({\bf
x}_1,...,{\bf x}_n)|0>
\l{bb4}
\ee
is looked for in the following form \c{Ber,MS-RJMP}
\be
\Psi_t = \sum_{n=0}^{\infty} \frac{1}{\sqrt{n!}} \int d{\bf  x}_1  ...
d{\bf x}_n   A_t^+({\bf   x}_1)  ...  A_t^+({\bf  x}_n)  \Psi_{0,n}  ({\bf
x}_1,...,{\bf x}_n)|0>_t
\l{bb5}
\ee
with
\be
|0>_t = c^t \exp [\frac{1}{2}\int d{\bf x}d{\bf  y}  M^t({\bf  x},{\bf
y}) A^+({\bf x}) A^+({\bf y})]|0>.
\l{bb6}
\ee
while operators $A^{+}_t({\bf x})$ are chosen to be
$$
A_t^{+}({\bf x}) = \int d{\bf y} [A^+({\bf y}) G_t^*({\bf y},{\bf  x})
- A^-({\bf y}) A_t^*({\bf y},{\bf x})].
$$
Namely, the Gaussian ansatz \r{bb6} formally satisfies eq.\r{bb3} if
\bea
i \frac{dc^t}{dt}  =   \frac{1}{2}   Tr   {\cal   H}^{--}_tM^t   c^t   +
\overline{\cal H}_t c^t,\crr
i \frac{dM^t}{dt} = {\cal H}^{++}_t + {\cal H}_t^{+-} M_t + M_t  {\cal
H}_t^{-+} + M_t {\cal H}_t^{--} M_t.
\l{bb7}
\eea
Here $M_t$ is the operator with kernel $M^t({\bf x},{\bf y})$,  ${\cal
H}_t^{-+} =  ({\cal  H}_t^{+-})^*$.  The  operators  $A_t^+({\bf  x})$
commute with $i\frac{d}{dt}-H_t$ if
\bea
i \frac{dF^t}{dt} = {\cal H}_t^{+-} F_t + {\cal H}_t^{++} G_t,
\qquad
-i \frac{dG^t}{dt} = {\cal H}_t^{-+} G_t + {\cal H}_t^{--} F_t.
\l{bb8}
\eea
Here $F_t$,  $G_t$  are  operators with kernels $F_t({\bf x},{\bf y})$
and $G_t({\bf x},{\bf y})$. Note that the operator $M_t = F_tG_t^{-1}$
formally satisfies   eq.\r{bb7}.   Initial   conditions   \r{bb4}  are
satisfied if $F_0=0$, $G_0=1$.

Let us check that eq.\r{bb3} is satisfied in a strong sense.

First of all, let us present some auxiliary lemmas.

{\bf Lemma 4.5.} {\it  Let  $M$  be  a  Hilbert-Schmidt  operator  and
$||M||<1$. Then
\be
\exp[\frac{1}{2} A^+MA^+]|0>
\l{bb8*}
\ee
}

The proof is presented in \c{Ber}.

{\bf Corollary.}  {\it  For  the  state   \r{bb8*},   the   estimation
\be
||\Psi_n|| \le  A  e^{- \alpha  n}
\l{bb9}
\ee
is  satisfied under conditions of
lemma 4.5 for some $A$ and $0< \alpha \le - \frac{1}{2}log ||M||$. }

{\bf Proof.} Since $||M||<1$, $||e^{2\alpha}M|| <1$. Since expression
$\tilde{\Psi} =  exp[\frac{1}{2}  e^{2\alpha} A^+ MA^+]|0>$ specifies a
Fock space   vector,   $||\tilde{\Psi}_{2n}||   =   ||e^{2\alpha   n}
\Psi_{2n}|| \le A$. Corollary is proved.

{\bf Lemma   4.6.}   {\it  Let  $M$,  $\delta  M$  be  Hilbert-Schmidt
operators, $||M||\le 1$, $||M+\delta M||\le 1$ and
$$
||\delta M||_2 \le \frac{1}{4} \log||M||^{-1} ||M||^{-3/8}.
$$
Then
$$
\exp[\frac{1}{2}A^+(M+\delta M)A^+]|0>      =      \sum_{k=0}^{\infty}
\frac{1}{k!} [\frac{1}{2}A^+\delta M A^+]^k \exp[\frac{1}{2}A^+MA^+]|0>
$$
}

{\bf Proof.} One should check that
\bea
s-\lim_{N\to\infty} \sum_{k,l,  k+l\le  N}  \frac{1}{2^kk!} (A^+\delta M
A^+)^k \frac{1}{2^ll!}    (A^+MA^+)^l    |0>    = \crr
s-\lim_{N\to\infty}
\sum_{k=0}^N \frac{1}{2^kk!}         (A^+\delta        M        A^+)^k
e^{\frac{1}{2}A^+MA^+}|0>
\l{bb10*}
\eea
Since the strong limit in the left-hand side of equality \r{bb10*}  exists,
eq.\r{bb10*} can be presemted as
\be
\sum_{k=0}^N \sum_{l=N-k+1}^{\infty} \Psi_{k,l} \to_{N\to\infty} 0
\l{bb10}
\ee
with
$$
\Psi_{k,l} =   \frac{1}{2^kk!}   (A^+\delta   MA^+)^k  \frac{1}{2^ll!}
(A^+MA^+)^l |0>.
$$
Since
$$
([A^+\delta M A^+]\Psi)_n({\bf x}_1,...,{\bf x}_n) = Sym \sqrt{n(n-1)}
\delta M({\bf x}_1,{\bf x}_2) \Psi_{n-2} ({\bf x}_3,...,{\bf x}_n),
$$
one has
$$
||([A^+\delta MA^+]\Psi)_n||    \le   \sqrt{n(n-1)}   ||\delta   M||_2
||\Psi||_{n-2}.
$$
By induction, one obtains:
$$
||[A^+\delta MA^+]^k   \Psi_{n-2k}||   \le   \sqrt{\frac{n!}{(n-2k)!}}
||\delta M||_2^k ||\Psi_{n-2k}||.
$$
It follows from the extimation \r{bb9} that
\beb
||\Psi_{k,l}|| \le \sqrt{\frac{(l+2k)!}{k!}} \frac{||\delta M||_2^k}
{2^kk!} A  e^{-\alpha  l/2}  e^{-\alpha  l/2} \crr
\le   \max_l   (l+2k)^k
e^{-\alpha (l+2k)/2}   A   e^{-\alpha   l/2}   \frac{(||\delta   M||_2
e^{\alpha})^k}{2^kk!} =  A    e^{-\alpha    l/2}    \frac{k^k}{k!e^k}
\left(\frac{||\delta M||_2e^{\alpha}}{\alpha}\right)^k
\eeb
Since $k!   \sim   (k/e)^k\sqrt{2\pi  k}$  as  $k\to\infty$,  one  has
$e^{-k}k^k/k!\le A_1$. Therefore,
\be
||\Psi_{k,l}|| \le AA_1 e^{-\alpha l/2} b^k
\l{bb10x}
\ee
with $b = ||\delta M||_2e^{\alpha}/\alpha$. Therefore,
$$
\sum_{k=0}^N \sum_{l=N-k+1}^{\infty}   ||\Psi_{k,l}||  =  \sum_{k=0}^N
AA_1 b^k e^{-\frac{\alpha}{2}(N-k+1)} \frac{1}{1-e^{-\alpha/2}}
\le AA_1 \frac{e^{-\alpha(N+1)/2}}{(1-e^{-\alpha/2})(1-be^{-\alpha/2})}.
$$
Therefore, for  $||\delta  M||_2  e^{3\alpha/2}  \le  \alpha$ property
\r{bb10} is satisfied. Choosing $\alpha = - \frac{1}{4} log ||M||$, we
obtain statement of lemma.

{\bf Lemma 4.7.} {\it Let $M_t$,  $t\in [t_1,t_2]$ be a differentiable
operator function, $||M_t||_2 < \infty$,
\be
|| \frac{M_{t+\delta   t}-M_t}{\delta   t}    -    \frac{dM_t}{dt}||_2
\to_{\delta t \to 0} 0.
\l{bb10a}
\ee
Then
\be
|| \frac{e^{\frac{1}{2}A^+M_{t+\delta t}A^+}|0>
- e^{\frac{1}{2}A^+M_{t}A^+}|0>}{\delta  t}  -  \frac{1}{2} A^+
\frac{dM_t}{dt} A^+e^{\frac{1}{2}A^+M_{t}A^+}|0>||_m
\to_{\delta t \to 0} 0.
\l{bb11}
\ee
}

{\bf Proof.}  Denote  $\delta  M  \equiv  \delta  M_{t,\delta   t}   =
M_{t+\delta t} - M_t$. It is sufficient to check the following formulas:
\be
||\frac{e^{\frac{1}{2}A^+M_{t+\delta t}A^+}|0> - 1 -
{\frac{1}{2}A^+M_{t}A^+}|0>}{\delta t}
e^{\frac{1}{2}A^+M_{t}A^+}|0>||_m \to_{\delta t \to 0} 0;
\l{bb12}
\ee
\be
||A^+\frac{\delta M}{\delta t}- \frac{dM}{dt})A^+
e^{\frac{1}{2}A^+M_{t}A^+}|0>||_m \to_{\delta t \to 0} 0.
\l{bb13}
\ee
The latter formula is a direct corollary of lemma 4.3, property
$||e^{\frac{1}{2}A^+M_{t+\delta t}A^+}|0>||_{m+1} <\infty$  following
from formula \r{bb9} and relation
$||\frac{\delta M}{\delta t} - \frac{dM}{dt}||_2 \to_{\delta t \to  0}
0$. Formula \r{bb12} is a corollary of the relation
\be
\sum_{k=2}^{\infty} \sum_{l=0}^{\infty}  \frac{1}{\delta t} (2k+1+l)^m
||\Psi_{k,l}|| \to_{\delta t \to 0} 0.
\l{bb14}
\ee
Makibg use   of   the   estimation  \r{bb10x}  and  formula  $||\delta
M||_2^2/\delta t \to_{\delta t\to 0} 0$,  we prove relation  \r{bb14}.
Lemma 4.7 is proved.

{\bf Lemma  4.8.}  {\it  Let  $T$ be such nonbounded self-adjoint
operator in $L^2({\bf R}^l)$
that $T-1 >0$, $D(T)\subset D({\cal H}^{+-})$, ${\cal H}_t^{+-}T^{-1}$
be uniformly bpinded operator.
Let the initial condition for eq.\r{bb3} be of
the form \r{bb4}, where $\Psi_{0,n}=0$ as $n\ge N_0$,
\be
\Psi_{0,n}({\bf x}_1,...,{\bf  x}_n)  =  \sum_{j=1}^{J_0}   f_j^1({\bf
x}_1) ... f_j^n({\bf x}_n), \qquad f_j^s \in D(T).
\l{bb14a}
\ee
Let Hilbert-Schmidt operator $M_t$ satisfy eq.\r{bb7}  (the  derivative
is defined   in   the  Hilbert-Schmidt  sense  \r{bb10a})  and  initial
condition $M_0=0$,  ,  $c_t$  obey  eq.\r{bb7},  $F_t$  and  $G_t$  be
uniformly bounded operators $F_t:D(T)\to D(T)$, $G_t:D(T)\to D(T)$
satisfying eq.\r{bb8} in the strong sense on $D(T)$, $F_0=0$, $G_0=1$.
Then the Fock vector \r{bb5} obeys eq.\r{bb3} in the strong sense.
}

{\bf Proof.} It is sufficient to prove lemma for the initial condition
$$
\Psi_0 = \frac{1}{\sqrt{n!}} A^+[f^1] ... A^+[f^n]|0>
$$
where $A^+[f]= \int d{\bf x} f({\bf x})A^+({\bf x})$. Let us show that
the Fock vector
$$
\Psi_t = \frac{1}{\sqrt{n!}} A_t^+[f^1] ... A_t^+[f^n]|0>_t
$$
with
$$
A_t^+[f] =  \int  d{\bf  y} [A^+({\bf y}) (G_t^*f)({\bf y}) - A^-({\bf
y}) (F_t^*f)({\bf y})]
$$
satisfies eq.\r{bb3}. Let
$$
\dot{\Psi}_t \equiv \frac{1}{\sqrt{n!}}  (\sum_{j=1}^n  A_t^+[f^1]  ...
\dot{A}_t^+[f^j] ... A_t^+[f^n]|0>_t + A_t^+[f^1]  ...
{A}_t^+[f^j] ... A_t^+[f^n]\frac{d}{dt}|0>_t
$$
with
\beb
\dot{A}_t^+[f] =  \int  d{\bf  y} [A^+({\bf y})
\frac{d}{dt}(G_t^*f)({\bf y}) - A^-({\bf
y}) \frac{d}{dt}(F_t^*f)({\bf y})], \crr
\frac{d}{dt}|0>_t \equiv \frac{dc^t}{dt} e^{\frac{1}{2} A^+M_tA^+}|0>
+ c^t \frac{1}{2} A^+ \frac{dM_t}{dt} A^+ e^{\frac{1}{2} A^+M_tA^+}|0>.
\eeb
One has
\beb
\frac{\Psi_{t+\delta t}-\Psi_t}{\delta    t}    -    \dot{\Psi}_t    =
\frac{1}{\sqrt{n!}} A^+_{t+\delta t} [f^1] ... A^+_{t+\delta t} [f^n]
\left(
\frac{|0>_{t+\delta t} - |0>_t}{\delta t} - \frac{d}{dt}|0>_t
\right)
+ \crr
\frac{1}{\sqrt{n!}} [ A^+_{t+\delta t} [f^1] ... A^+_{t+\delta t} [f^n]
-  A^+_{t} [f^1] ... A^+_{t} [f^n]] \frac{d}{dt}|0>_t +\crr
\sum_{j=1}^n  A^+_{t+\delta t} [f^1] ... A^+_{t+\delta t} [f^{j-1}]
[\frac{A^+_{t+\delta t}[f^j]     -     A^+_t[f^j]}{\delta     t}     -
\dot{A}^+_t[f^j]]  A^+_{t} [f^{j+1}] ... A^+_{t} [f^n]|0>_t +\crr
\sum_{j=1}^n \frac{1}{\sqrt{n!}}
(A^+_{t+\delta t}[f^1]...A_{t+\delta t}^+[f^{j-1}] -
 A^+_{t} [f^1] ... A^+_{t} [f^{j-1}]] \dot{A}_t^+[f^j]
 A^+_{t} [f^{j+1}] ... A^+_{t} [f^n]|0>_t.
\eeb
It follows from lemmas 4.3, 4.7 and conditions of lemma 4.8 that
$$
||\frac{\Psi_{t+\delta t}-   \Psi_t}{\delta   t}   -    \dot{\Psi}_t||
\to_{\delta t\to 0} 0.
$$
Eqs.\r{bb7}, \r{bb8}  imply  that $\dot{\Psi}_t = -iH_t\Psi_t$.  Lemma
4.8 is proved.

Denote by ${\cal D}_1 \subset {\cal F}$ the set of  all  Fock  vectors
$\Psi \in  {\cal  F}$ such that $\Psi_n$ vanish at $n\ge N_0$ and have
the form \r{bb14a} as $n<N_0$.  Lemma 4.8 allows us to  construct  the
mapping $U_t:{\cal  D}_1 \to {\cal F}$ of the form $U_t\Psi_0=\Psi_t$.
Note that the domain ${\cal D}_1$ is dense in ${\cal F}$.

Denote
$$
A_t^-[f] \equiv  (A_t^+[f])^+  \equiv  \int  d{\bf  y}  [A^-({\bf  y})
(G_tf)({\bf y}) - A^+({\bf y}) (F_tf)({\bf y}) ].
$$

{\bf Lemma 4.9.} {\it
1. The operators $A_t^{\pm}[f]$ obey the commutation relations
\be
[A^-_t[f], A^+_t[g]] = (f,g), \qquad [A^{\pm}_t[f], A^{\pm}_t[g]] =0.
\l{bb15}
\ee
2. The following property is satisfied:
\be
A_t^-[f]|0>_t =0.
\l{bb15a}
\ee
3. The operator $U_t$ is isometric.
}

{\bf Proof.} The commutation relations \r{bb15} are rewritten as
\bea
(G_tf,G_tg) - (F_tf,F_tg) = (f,g);\crr
(F_t^*f, G_tg) - (G_t^*f,F_tg) =0.
\l{bb16}
\eea
They are satisfied at $t=0$.  The time derivatives  of  the  left-hand
sides of  eqs.\r{bb16}  vanish because of eqs.\r{bb8}.  Statement 1 is
proved.

The fact that $U_t$ is an isometric operator is  a  corollary  of  the
property $\frac{d}{dt}(\Psi_t,\Psi_t) =0$.

Analogously to  lemma  4.8,  we find that the vector $\tilde{\Psi}_t =
A^-_t[f]|0>_t$ obeys eq.\r{bb3} in the strong sense.  Since $\Psi_0=0$
and $||\Psi_t||=||\Psi_0||$, one has $\Psi_t=0$. Property \r{bb15a} is
proved. Note that it means that
\bea
M_tG_t=F_t.
\l{bb16c}
\eea
Lemma 4.9 is proved.

Therefore, the operator $U_t$ can be extended to the whole space $\cal
F$, $U_t:{\cal F} \to {\cal F}$.

{\bf Lemma 4.10.} {\it Let the operator
\beb
\left(
\matrix{
G^+ & - F^+ \crr - F^T & G^T
}
\right)
\eeb
be invertible.  Then the following relation  is  satisfied  on  ${\cal
D}_1$:
\be
U_t^{-1} A^+TA^- U_t \Psi_0= (A^+G^T_t + A^-F^+)T(FA^++G^*A^-)\Psi_0
\l{bb16a}
\ee
}

{\bf Proof.} It follows from lemma 4.9 that
\beb
\left(
\matrix{
G^+ & - F^+ \crr - F^T & G^T
}
\right)
\left(
\matrix{
G &  F^* \crr  F & G^*
}
\right) = 1
\eeb
Therefore,
\beb
\left(
\matrix{
G^+ & - F^+ \crr - F^T & G^T
}
\right)^{-1} =
\left(
\matrix{
G &  F^* \crr  F & G^*
}
\right)
\eeb
and
\beb
A^-({\bf y}) = \int d{\bf z} (F_t({\bf y},{\bf z})A_t^+({\bf z}) +
G_t^*({\bf y},{\bf z}) A_t^-({\bf z})), \crr
A^+({\bf y}) = \int d{\bf z} (F^*_t({\bf y},{\bf z})A_t^-({\bf z}) +
G_t({\bf y},{\bf z}) A_t^+({\bf z})).
\eeb
Identity \r{bb16a} is then a corollary of definition of  the  operator
$U_t$.

{\bf Lemma 4.11.} {\it
Let $\Psi_0 \in {\cal D}$.  Suppose that $TF_t$  and
${\cal H}^{++}$ are
continuous operator   functions   in   the  $||\cdot||_2$-norm,  $G_t$,
$T^{1/2}G_tT^{-1/2}$, $TG_tT^{-1}$, $T^{-1/2}{\cal H}_t^{+-}T^{-1/2}$,
${\cal H}^{+-}T^{-1}$  are  continous  operator   functions   in   the
$||\cdot||$-norm. Then the following statements are satisfied.

1. $\Psi_t \equiv U_t\Psi_0 \in {\cal D}$.

2. $\Psi_t$ obeys eq.\r{bb3} in the strong sense.

3.
\be
||\Psi_t-\Psi_0||_1^T \to_{t\to 0} 0.
\l{bb17}
\ee
}

{\bf Proof.} Let $\Psi_0\in {\cal D}_1$.  For $||U_t\Psi_0||_1^T$, one
has the following estimation:
\beb
||U_t\Psi_0||_1^T =  ||U_t^{-1}(\hat{T}+1)U_t\Psi_0|| \le ||\Psi_0|| +
||(A^+G^T+A^-F^+)T(FA^++G^*A^-)\Psi_0|| \le
\crr
(1 +  ||F^+||_2 ||TF||_2)
||\Psi_0||  +  (\sqrt{2}||G^TTF||_2  + ||F^+TF|| +
||F^+TG||_2) ||\Psi_0|| \crr
+ (||T^{-1/2}G^TTG^*T^{-1/2}||  + ||A^TTA^*T^{-1}||)||\Psi_0||_1^T \le
const ||\Psi||_1^T
\eeb
at $t\in[0,t_1]$.  Therefore,the  operator  $U_t$  is  bounded in norm
$||\cdot||_1^T$. The extension of the operator $U_t$ to  $\cal  D$  is
then also  a  bounded operator in $||\cdot||_1^T$ norm.  One therefore
has $\Psi_t\in {\cal D}$.

The fact that $||U_t\Psi_0 - \Psi_0||_1^T \to_{t\to 0} 0$  if  $\Psi_0
\in {\cal  D}_1$  is  justified  analogously  to lemma 4.8.  Since the
operator $U_t:{\cal D} \to {\cal D}$ is  uniformly  bounded  at  $t\in
[0,t_1]$ in  $||\cdot||_1^T$-norm,  the Banach-Steinhaus theorem (see,
for example, \c{KA}) implies relation \r{bb17}.

To check the second statement, note that lemma 4.8 imply that
\be
\frac{U_{t+\delta t}-U_t}{\delta t} \to_{\delta t \to 0} \frac{dU_t}{dt}
\l{bb18}
\ee
in the  strong  sense  on  ${\cal  D}_1$.  For  showing  that relation
\r{bb18} is  satisfied  in  the  strong  sense  on  $\cal  D$,  it  is
sufficient to show that the operator
$$
\frac{\delta U}{\delta t} :{\cal D} \to {\cal F}
$$
is uniformly bounded,
$$
||\frac{\delta U}{\delta t} \Psi|| \le C ||\Psi||_1^T.
$$
One has
\beb
||\frac{\delta U}{\delta t} \Psi|| = ||\int_0^1 ds  \dot{U}_{t+s\delta
t}\Psi|| = ||\int_0^1 ds H_{t+s\delta t} U_{t+s\delta t}|| \le
\crr
\max_{s\in[0,1]} [\sqrt{2}||{\cal     H}^{++}_{t+s\delta    t}||_2    +
||T^{-1/2}{\cal H}^{+-}_{t_s+ \delta t}T^{-1/2}|| +
||{\cal H}^{+-}_{t_s+\delta t}T^{-1}|| ] ||U_{t+s\delta t}\Psi||_1^T.
\eeb
Lemma 4.11 is proved.

Let us now check properties of operators $F_t$, $F_t$, $M_t$.

First of all, consider the Cauchy problem
\bea
i\dot{f}_t = Y_tf_t + Z_tg_t,\crr
-i\dot{g}_t = Z_t^*f_t + Y_t^*g_t,\crr
f_0=0, g_0=1,
\l{bb19}
\eea
where $g_t$  is   a   bounded   operator   functions,   $f_t$   is   a
Hilbert-Schmidt operator  function.  The  derivatives  in \r{bb19} are
understood as
\be
||(\frac{g_{t+\delta t}-g_t}{\delta    t}    -     \dot{g}_t)\varphi||
\to_{\delta t \to 0} 0, \qquad
||\frac{f_{t+\delta t}-f_t}{\delta    t}    -     \dot{f}_t||_2
\to_{\delta t \to 0} 0.
\l{bb19*}
\ee

{\bf Lemma  4.12.}  {\it  Let  $Y_t$  be a strongly continous operator
function, while $||Z_{t+\tau}-Z_t||_2 \to_{\tau\to 0} 0$,  $||TZ_t||_2
\le a_1^t$,  $||TY_tT^{-1}||  \le  a_2^t$,  $||T^{1/2}Y_tT^{-1}||  \le
a_3^t$ for smooth functions $a_k^t$.  Then there exist a  solution  to
the Cauchy problem \r{bb19} such that
\be
||Tf_t||_2 \le a_4^t, \qquad ||T^{1/2}g_tT^{-1/2}|| \le a_5^t, \qquad
||Tg_tT^{-1}|| \le a_6^t, \qquad ||g_t|| \le a_7^t
\l{bb19a}
\ee
for smooth functions $a_k^t$.
}

{\bf Proof} (cf.\c{MS-RJMP}). Let us look for the solution to the
Cauchy problem in the following form:
\be
f_t = \sum_{n=0}^{\infty} f_t^n, \qquad
g_t = \sum_{n=0}^{\infty} g_t^n.
\l{bb20}
\ee
where $f_t^0=0$, $g_t^0=1$,
\bea
f_t^{n+1} =  -i  \int_0^t   d\tau   (Y_{\tau}f_{\tau}^n   +   Z_{\tau}
g_{\tau}^n), \crr
g_t^{n+1} = -i \int_0^t d\tau (Y^*_{\tau}g_{\tau}^n + Z^*_{\tau}
f_{\tau}^n).
\l{bb20*}
\eea
By induction we find that $||f_t^n||_2 \le C_1t^n/n!$,
$||g_t||_2 \le C_1t^n/n!$ for $t\in [0,t_1]$. Here $C_1$ is a constant.

Therefore, the series \r{bb20} converge.  $f_t$  is  a  Hilbert-Schmidt
operator, while $g_t$ is a bounded operator. Analogously, we show
$$
||Tf_t^n||_2 \le \frac{C_2 t^n}{n!},
\qquad
||Tg_t^nT^{-1}|| \le \frac{C_2 t^n}{n!},
\qquad
||T^{1/2}g_t^n T^{-1/2}|| \le \frac{C_2 t^n}{n!},
$$
where $t\in [0,t_1]$. Therefore, properties \r{bb19a} are satisfied.

To check relations \r{bb19*}, note that
\bea
f_t = - i \int_0^t d\tau (Y_{\tau}f_{\tau} + Z_{\tau}g_{\tau}),
\crr
g_t = i\int_0^t d\tau (Z_{\tau}^*f_{\tau} + Y_{\tau}^*g_{\tau}).
\l{bb20a}
\eea
Eqs.\r{bb20a} imply that the  operator  functions  $f_t$,  $g_t$  obey
properties
$$
||T(f_{t+\delta t} - f_t)||_2 \to_{\delta t \to 0} 0,
\qquad
||(g_{t+\delta t} - g_t)|| \to_{\delta t \to 0} 0,
$$
Therefore,
\beb
||i \frac{f_{t+\delta t} - f_t}{\delta t} - Y_t f_t - Z_t g_t||_2 \le
\int_0^1 ds ||Y_{t+s\delta t} f_{t+s\delta t} + Z_{t+s\delta t} g_{t+s
\delta t} - Y_tf_t - Z_tg_t||_2, \crr
||(-i \frac{g_{t+\delta  t}  -  g_t}{\delta  t}  -  Z_t^*f_t  -  Y_t^*
g_t)\varphi_t|| \le
\int_0^1 ds  ||(Z_{t+s\delta  t}^* f_{t+s\delta t} + Y^*_{t+s\delta t}
g_{t+s\delta t} - Z_t^* f_t - Y_t^* g_t) \varphi_t||.
\eeb
Since the  integrands  are  uniformly bounded functions,  the Lesbegue
theorem (see,  for example,  \c{KF}) tells us that it is sufficient to
check that
\beb
||Y_{t+\tau}f_{t+\tau} - Y_tf_t||_2 \to_{\delta t\to 0} 0,
\qquad
||Z_{t+\tau}g_{t+\tau} - Z_tg_t||_2 \to_{\delta t\to 0} 0,\crr
s-\lim_{\tau\to 0} Z_{t+\tau}^* f_{t+\tau} = Z_t^*f_t,\crr
s-\lim_{\tau\to 0} Y_{t+\tau}^* g_{t+\tau} = Y_t^*g_t.
\eeb
These relations  are  corollaries  of  conditions  of  lemma  4.12 and
formulas \r{bb20*}.

{\bf Lemma 4.13.} {\it Let ${\cal H}_t^{+-} = L + {\cal H}_t$,  ${\cal
H}_t$, $T^{1/2}{\cal   H}T^{-1/2}$,   $T{\cal  H}T^{-1}$  be  strongly
continous operator functions, $||{\cal H}^{++}_{t+\delta t} -
{\cal H}^{++}_{t}||_2   \to_{\delta   t   \to   0}   0$,   $L$   be  a
$t$-independent (maybe nonbounded) self-adjoint  operator,  such  that
$||LT^{-1}||<\infty$, while  $||T^{1/2}  e^{-iLt}  T^{-1/2}||<\infty$,
$||T e^{-iLt} T^{-1}|| <\infty$.  Then there exists a solution to  the
Cauchy problem  for  system \r{bb8} for the initial condition $F_0=0$,
$G_0=1$:
\bea
||i \frac{F_{t+\delta t} - F_t}{\delta t} - {\cal H}_t^{++}
F_t - {\cal H}_t^{+-} G_t||_2 \to_{\delta t\to 0} 0,\crr
||(-i \frac{G_{t+\delta t} - G_t}{\delta t} - {\cal H}_t^{--}
G_t - {\cal H}_t^{-+} F_t)\varphi||  \to_{\delta  t\to  0}  0,  \qquad
\varphi \in D(T).
\l{bb22}
\eea
Moreover,
\be
||TF_t||_2 \le b(t), \qquad
||TG_tT^{-1}|| \le b(t), \qquad
||T^{1/2} G_t T^{-1/2}|| \le b(t), \qquad
||G_t|| \le b(t)
\l{bb21}
\ee
for some smooth function $b(t)$ on $t\in [0,t_1]$.
The properties \r{bb16c} are also satisfied.
}

{\bf Proof.} Consider the operator functions
$$
F_t = e^{-iLt} f_t, \qquad G_t = e^{iL^*t} g_t,
$$
where $(f_t, g_t)$ is a solution to the Cauchy problem \r{bb19} with
$Y_t =  e^{iLt}  {\cal H}_t e^{-iLt}$,  $Z_t = e^{iLt} {\cal H}_t^{++}
e^{iL^*t}$, $f_0=0$,   $g_0=1$.   Check   of   properties   \r{bb21}  is
straightforward. Let us prove relations \r{bb22}. One has
\beb
i \frac{F_{t+\delta  t}-F_t}{\delta  t}  -  (L+{\cal  H}_t)F_t - {\cal
H}_t^{++} G_t = (i \frac{e^{-iL\delta t} - 1}{\delta t }T^{-1} -LT^{-1})
TF_t + i e^{-iLt} (\frac{f_{t+\delta  t}-f_t}{\delta  t}-  \dot{f}_t),
\crr
- i \frac{G_{t+\delta  t}-G_t}{\delta  t}  -  (L^*+{\cal  H}^*_t)G_t - {\cal
H}_t^{--} F_t = (-i \frac{e^{-iL^*\delta t} - 1}{\delta t }T^{-1} -L^*T^{-1})
TF_t + i e^{-iL^*t} (\frac{g_{t+\delta t}-g_t}{\delta t}- \dot{g}_t).
\eeb
Since
$$
||(i\frac{e^{-iL\tau}-1}{\tau}T^{-1} - LT^{-1})\varphi||
\le \int_0^1 ds ||(e^{-iL\tau s}-1) LT^{-1}\varphi||\to_{\tau\to 0} 0,
$$
we obtain relations \r{bb22}.

Property \r{bb16c} is proved analogously  to  \c{MS-RJMP}:  one  should
consider the convergent in $||\cdot||$-norm series
\beb
\left(
\matrix{
G & F^* \crr F & G^*
}
\right)^{-1} =
\sum_{n=0}^{\infty}
\left(
\matrix{
G_t^{(-n)} & F_t^{(-n)*} \crr F_t^{(-n)} & G_t^{(-n)*}
}
\right)
\left(
\matrix{
e^{iL^*t} & 0 \crr 0 & e^{-iLt}
}
\right)
\eeb
with
\beb
\left(
\matrix{
G_t^{(-n)} & F_t^{(-n)*} \crr F_t^{(-n)} & G_t^{(-n)*}
}
\right)
= i \int_0^t d\tau
\left(
\matrix{
G_{\tau}^{(-n+1)} & F_{\tau}^{(-n+1)*} \crr F_{\tau}^{(-n+1)} &
G_{\tau}^{(-n+1)*}
}
\right)
\left(
\matrix{
Y_{\tau} & Z_{\tau} \crr -Z^*_{\tau} & -Y^*_{\tau}
}
\right)
\eeb
Lemma 4.13 is proved.

{\bf Lemma  4.14.}  {\it Under conditions of lemma 4.13 there exists a
solution to  the  Cauchy  problem  for  eq.\r{bb7}  with  the  initial
condition $M_0=0$.}

{\bf Proof.} It follows from \c{Ber} that the matrix $G$ is invertible
and $||G^{-1}||<1$. Consider the operator $M_t=F_tG_t^{-1}$. Note that
$||TM_t||_2 < \infty$, $||LM_t|| <\infty$. One has
$$
M_{t+\delta t}  - M_t = M_{t+\delta t} (G_t-G_{t+\delta t}) G_t^{-1} +
(F_{t+\delta t} - F_t) G_t^{-1},
$$
so that  $||T(M_{t+\delta  t}  -  M_t)||_2  \to_{\delta  t\to  0}  0$.
Therefore,
\beb
\frac{M_{t+\delta t}-M_t}{\delta t} - \dot{F}_tG_t^{-1} + F_T G_t^{-1}
\dot{G}_t G_t^{-1} = \crr
M_{t+\delta t}T T^{-1} (\frac{G_t-G_{t+\delta t}}{\delta t} - \dot{G}_t)
G_t^{-1} +   (M_{t+\delta   t}  -  M_t)T  T^{-1}\dot{G}_t  G_t^{-1}  +
(\frac{F_{t+\delta t}-F_t}{\delta t} - \dot{F}_t) G_t^{-1}.
\eeb
Ananlogously to lemmas 4.12, 4.13, one finds
$$
||(\frac{G^+_{t+\delta t}          -G^+_t}{\delta         t}         -
\dot{G}_t^+)T^{-1}\varphi|| \to_{\delta t \to 0} 0.
$$
Therefore,
$$
||\frac{M_{t+\delta t}-M_t}{\delta t}- \dot{M}_t||_2
\to_{\delta  t  \to 0} 0.
$$
Lemma 4.14 is proved.

Therefore, we have proved the following theorem.

{\bf Lemma  4.15.}   {\it  Let  $T$,  $L$  be
self-adjoint operators in $L^2({\bf R}^l)$ such that
\beb
||T^{-1/2}LT^{-1/2}|| <\infty, \qquad
||LT^{-1}|| <\infty, \qquad ||T^{1/2}e^{-iLt}T^{-1/2}|| \le C,
\crr
||Te^{-iLt}T^{-1}|| \le C, \qquad t\in [0,t_1].
\eeb
Let $T-c$ be positively definite for some positive constant $c$,
${\cal   H}^{+-}_t   =   L+{\cal   H}_t$,   ${\cal   H}^{++}$   be
operator-valued functions such that
$||T({\cal H}^{++}_{t+\delta t} - {\cal H}^{++}_t)||_2 \to_{\delta t\to
0} 0$,  ${\cal H}_t$, $T{\cal H}_tT^{-1}$, $T^{1/2}{\cal H}_tT^{-1/2}$
are strongly  continous  operator  functions,  $\overline{H}_t$  be  a
continous function.  Then there exists a unique solution  $\Psi_t$  to
the Cauchy problem \r{bb3},  provided that $\Psi_0 \in {\cal D} \equiv
\{\Psi \in  {\cal  F}  |  ||\Psi||_1^T  <\infty\}$  It  satisfies  the
properties $\Psi_t    \in   {\cal   D}$   and   $||\Psi_t-\Psi_0||_1^T
\to_{\delta \to 0} 0$.
}

Thus, properties G1-G6 are obtained as corollaries of F1-F3.

\section{Composed semiclassical states}

We have already  mentioned  that  composed  semiclassical  states  are
specified by a set
$
\left(
\begin{array}{c}
X(\alpha) \\
g(\alpha) \in {\cal F}_{X(\alpha)}
\end{array}
\right)
$; $\alpha\in {\Lambda}^k$. The inner product is given by eq.\r{1.22}.
For the case ${\cal F}_{X(\alpha)}  =  {\cal  F}(L^2({\bf  R}^l)$  and
$\omega$ of the form \r{4.1}, expression \r{1.22} takes the form
\be
\int d\alpha (g(\alpha), \int d\beta
e^{\beta_a (A^+B_a - A^-B_a^*)} g(\alpha))
\l{5.1}
\ee
with
\be
B_s(\alpha,\cdot) = \varphi_X[\frac{\partial X}{\partial \alpha_s}].
\l{5.2}
\ee
Let us investigate the inner product space of composed states in  more
details.

\subsection{Constrained Fock space}

The purpose of this subsection is to investigate the properties of the
inner product
\be
<Y_1,Y_2> =
\int d\beta
(Y_1, exp  [\sum_{s=1}^k
\beta_s \int d{\bf x} (B_s({\bf x}) A^+({\bf
x}) - B_s^*({\bf x}) A_s^-({\bf x}))] Y_2)
\l{w18}
\ee
for the Fock vectors $Y_1$, $Y_2$. Suppose the functions $B_1,...,B_k$
to be linearly independent.  Since the inner product \r{w18} resembles
the inner  products for constrained systems \c{Marolf},  we will call the
space under construction as a constrained Fock space.

First of all,  investigate the problem of convergence of the  integral
\r{w18}. Note that the operator
$$
U[B] =  \exp  [\int  d{\bf  x} (B({\bf x}) A^+({\bf x}) - B^*({\bf x})
A^-({\bf x}))]
$$
is a well-defined unitary  operator  \c{Ber},  provided  that  $B  \in
L^2({\bf R}^l)$, and obey the relations
\beb
A^-({\bf x}) U[B] = U[B] (A^-({\bf x}) + B({\bf x}));\crr
A^+({\bf x}) U[B] = U[B] (A^+({\bf x}) + B^*({\bf x})).
\eeb

{\bf Lemma 5.1.} (cf. \c{MS3}). {\it The following estimation is satisfied:
\be
||B||^m |(Y_1,   U[B]  Y_2)|  \le  \sum_{k=0}^m  \frac{m!}{k!  (m-k)!}
||Y_1||_{k/2} ||Y_2||_{(m-k)/2}.
\l{w19}
\ee
}

{\bf Proof.} One has
$$
[\int d{\bf x} B^*({\bf x}) A^-({\bf x}); U[B]] = ||B||^2 U[B],
$$
so that
$$
||B||^2 (Y_1,  U[B] Y_2) = (\int d{\bf x} B({\bf x}) A^+({\bf x}) Y_1,
U[B] Y_2) - (Y_1, U[B] \int d{\bf x} B^*({\bf x}) A^-({\bf x}) Y_2).
$$
Applying this identity $m$ times, we obtain:
\beb
||B||^{2m} (Y_1,U[B]Y_2)  =   \sum_{k=0}^m   (-1)^{m-k}   \frac{m!}{k!
(m-k)!} ((\int  d{\bf  x}  B({\bf x}) A^+({\bf x}))^k Y_1,
\crr
U[B] (\int
d{\bf x} B^*({\bf x}) A^-({\bf x}))^{m-k} Y_2).
\eeb
Making use of the result of lemma 4.3,
$$
||\int d{\bf  x}  B({\bf  x})  A^{\pm}({\bf  x})   Y||_l   \le   ||B||
||Y||_{l+1/2},
$$
we find:
$$
||B||^{2m} |(Y_1,  U[B]  Y_2)|  \le  \sum_{s=0}^m  \frac{m!}{s!(m-s)!}
||B||^m ||Y_1||_{s/2} ||Y_2||_{(m-s)/2}.
$$
Lemma 5.1 is proved.

{\bf Corollary 1.} {\it  Let  $B_1,...,B_k$  be  linearly  independent
functions. Then  for some constant $C_1>0$ the following estimation is
satisfied:
$$
|\beta|^m |(Y_1,  U[\sum_s \beta_s B_s] Y_2)| \le C_1^m  ||Y_1||_{m/2}
||Y_2||_{m/2}.
$$
}

{\bf Proof.}  It is sufficient to notice that for linearly independent
$B_1,...,B_k$ the  matrix  $(B_m,B_s)$  is  not  degenerate,  so  that
$||\frac{1}{2} \sum_m \beta_m B_m||^2 \ge C_1^{-1} |\beta|^2$
for some $C_1$.  Applying the property $||Y||_{s/2}  \le  ||Y||_{m/2}$
for $s\le m$, making use of eq.\r{w19}, we prove corollary 1.

{\bf Corollary 2.} {\it Let $||Y_1||_{m/2} <\infty$,
$||Y_2||_{m/2} <\infty$ for some $m>k$, Then the integrand entering to
eq.\r{w18} obeys the relation
\be
|(Y_1, U[\sum_s\beta_s B_s] Y_2)| \le \frac{const}{(|\beta|+1)^m}
\l{w20}
\ee
and integral \r{w18} converges.
}

{\bf Corollary 3.} {\it Let $Y_{1,n}$,  $Y_{2,n}$ be such sequences of
Fock vectors    that     $||Y_{1,n}||_{m/2}     \to_{n\to\infty}     0$,
$||Y_{2,n}||_{m/2} \le    C$    for    some   $m>k$.   Then   $<Y_1,Y_2>
\to_{n\to\infty} 0$. }

Let us investigate the property of  nonnegative  definiteness  of  the
inner product \r{w18}.

{\bf Lemma  5.2.}  {\it  Let  $||Y||_m <\infty$ for some $m>k$ and $Im
(B_s,B_l)=0$. Then $<Y,Y> \ge 0$.
}

{\bf Proof.} Introduce the following "regularized" inner  product
$$
<Y,Y>_{\varepsilon} =  \int  d\beta  e^{-{\varepsilon}  |\beta|^2} (Y,
U[\sum_s \beta_s B_s] Y).
$$
It follows form estimation \r{w20} and  the  Lesbegue  theorem  \c{KF}
that
$$
<Y,Y>_{\varepsilon} \to_{{\varepsilon}\to 0} 0.
$$
It is  sufficient then to prove that $<Y,Y>_{\varepsilon} \ge 0$.  One
has:
$$
e^{-{\varepsilon}|\beta|^2} = (4\pi{\varepsilon})^{k/2}  \int  d\beta'
e^{-2{\varepsilon}|\beta-\beta'|^2 - 2{\varepsilon}|\beta'|^2}
$$
Therefore,
\be
<Y,Y>_{\varepsilon} =  \int d\beta' d\beta'' (4\pi{\varepsilon})^{k/2}
e^{-2{\varepsilon} (|\beta'|^2 + |\beta''|^2} (U[\sum_s \beta_s'' B_s]
Y, U[\sum_s \beta_s' B_s] Y),
\l{w21}
\ee
here the shift of variable $\beta = \beta' - \beta''$ is made. We have
also taken into account that
$$
U[\sum_s\beta_s' B_s] U[-\sum_s\beta_s'' B_s]
= U[\sum_s(\beta_s'-\beta_s'') B_s],
$$
provided that the operators
$$
\int d{\bf x} (B_s({\bf x})A^+({\bf x}) - B_s^*({\bf x}) A^-({\bf x})]
$$
commute (i.e. $Im(B_s,B_l)=0$).
Formula \r{w21} is taken to the form
$$
<Y,Y>_{\varepsilon} =    ||\int    d\beta    (4\pi{\varepsilon})^{k/4}
e^{-2{\varepsilon}|\beta|^2} U[\sum_s \beta_s B_s] Y||^2 \ge 0.
$$
Lemma 5.2 is proved.

The expression  \r{w18}  depends  on  $k$   functions   $B_1,...,B_k$.
However, one may perform linear substitutions of variables $\beta$, so
that only the subspace $span \{B_1, ..., B_k\}$ is essential.

{\bf Definition  5.1.}  {\it  A  $k$-dimensional  subspace  $L_k   \in
L^2({\bf R}^l)$  is  called as a $k$-dimensional isotropic plane if $Im
(B',B'') = 0$ for all $B',B'' \in L_k$.}

Let $L_k$ be a $k$-dimensional isotropic plane with an invariant under
shifts measure $d\sigma$.  Let $B_1,...,B_k$ be a basis on $L_k$.  One
can assign then coordinates $\beta_1,...,\beta_n$ to any element $B\in
L_k$ according  to  the  formula $B= \sum_s \beta_s B_s$.  The measure
$d\sigma$ is presented as $d\sigma = a d\beta_1 ... d\beta_k$ for some
constant $a$. Consider the inner product
\be
<Y_1,Y_2>_{L_k} =  a  \int  d\beta (Y_1,  U[\sum_s \beta_s B_s] Y_2) =
\int d\sigma (Y_1, U[B] Y_2),
\l{w22}
\ee
$||Y_{1,2}||_{[k/2+1]} \le \infty$. This definition is invariant under
change of basis.

By ${\cal F}_{[k/2+1]}$ we denote space of such Fock vectors $Y$  that
$||Y||_{[k/2+1]} <  \infty$.  We say that $Y\stackrel{L_k}{\sim} 0$ if
$<Y,Y>_{L_k} = 0$.  Thus,  the space ${\cal F}_{[k/2+1]}$  is  divided
into equivalence classes. Introduce the following inner product on the
factor-space ${\cal F}_{[k/2+1]}/\sim$:
\be
<[Y_1], [Y_2]>^{L_k} = <Y_1,Y_2>_{L_k}
\l{w23}
\ee
for all $Y_1 \in [Y_1]$,  $Y_2 \in [Y_2]$.  This definition is correct
because of the following statement.

{\bf Lemma 5.3.} {\it Let $<Y,Y>_{L_k} = 0$.  Then $<Y,Y'>_{L_k} =  0$
for all $Y'$.}

The proof is standard (cf,, forexample, \c{KF}). One has
$$
0 \le  <Y'  + \sigma Y,  Y'+\sigma Y>_{L_k} = <Y',Y'>_{L_k} + \sigma^*
<Y,Y'>_{L_k} + \sigma <Y',Y>_{L_k}
$$
for all $\sigma \in {\bf C}$, so that $<Y,Y'>_{L_k} = 0$.

{\bf Definition 5.2.} {\it A constrained  Fock  space  ${\cal  F}(L_k,
d\sigma)$ is the completeness of the factor-space
${\cal F}_{[k/2+1]}/\sim$ with respect to the inner product \r{w23},
$$
{\cal F}(L_k) = \overline{ {\cal F}_{[k/2+1]}/\sim}.
$$
}

\subsection{Transformations of constrained Fock vectors}

Let us investigate evolution of constrained Fock vectors. Consider the
Cauchy problem for eq.\r{bb3}. Denote
${\cal F}_m = \{ \Psi \in {\cal F} | ||\Psi||_m < \infty$.

{\bf Lemma 5.4.} {\it Let $\Psi_0 \in {\cal F}_m$.  Then  $\Psi_t  \in
{\cal F}_m$.}

{\bf Proof.} Analogously to proof of lemma 4.11, one has
\beb
||U_t\Psi_0|| = ||U_t^{-1} (A^+A^-+1)^m U_t \Psi_0|| = ||(1+ (A^+G_t^T +
A^-F_t^+)(F_t A^+ + G_t^* A^-))^m \Psi_0||
\eeb
It follows from Lemmas 4.2, 4.3 that
\beb
||(1+ (A^+G_t^T +
A^-F_t^+)(F_t A^+ + G_t^* A^-)) \Psi||_l \le \crr
||\Psi||_l +   ||F_t||_2^2   ||\Psi||_l  +  \sqrt{2}  ||G^T_t  F_t||_2
||\Psi||_{l+1} +  ||F^T_t  F_t^*  +  G^T_t  G^*_t||  ||\Psi||_{l+1}  +
||F^+_t G^*_t||_2 ||\Psi||_{l+1}\crr
\le C ||\Psi||_{l+1}
\eeb
with
$$
C = 1 +   ||F_t||_2^2  +  \sqrt{2}  ||G^T_t  F_t||_2
 +  ||F^T_t  F_t^*  +  G^T_t  G^*_t||  +
||F^+_t G^*_t||_2.
$$
Applying this estimation, we obtain by induction:
$$
||U_t\Psi_0||_m \le C^m ||\Psi_0||_m.
$$
Lemma is proved.

Let $L_k$ be a $k$-dimensional isotropic plane with invariant  measure
$d\sigma$. Define its evolution transformation $L_k^t$ as follows. Let
$(B_1,...,B_k)$ be a basis on $L_k$.  Let $B_s^t$ be solutions to  the
Cauchy problems
\bea
i \dot{B}_s^t = {\cal H}_t^{+-} B_s^t + {\cal H}_t^{++} (B_s^t)^*;
\crr
- i \dot{B}_s^{t*}  =  {\cal  H}_t^{-+}  B_s^{t*}  +  {\cal  H}_t^{--}
B_s^t;\crr
B_s^0 = B_s; B_s^{0*} = B_s^*.
\l{w23a}
\eea
they can be expressed as
\bea
B_s^t = F_t B_s^* + G_t^* B_s; \crr
B_s^{t*} = F_t^* B_s + G_t B_s^*.
\l{w23b}
\eea

{\bf Lemma  5.5.} {\it Let $Im (B_i,B_j)=0$.  Then $Im (B_i^t,B_j^t) =
0$.}

{\bf Proof.} One has:
\beb
2i Im(B_i^t,B_j^t) = (B_i^t,B_j^t) - (B_j^t,B_i^t) = \crr
(F_t B_i^* + G_t B_i, F_t B_j^* + G_t B_j) -
(F_t B_j^* + G_t B_j, F_t B_i^* + G_t B_i) = \crr
(B_i,B_j) - (B_j,B_i) = 0.
\eeb
because of relations \r{bb16} of Appendix B.

Therefore, $L_k^t$  is  also  an  isotropic plane.  Define the measure
$d\sigma^t$ on $L_k^t$ as  follows.  For  the  choice  of  coordinates
$\beta_1,..,\beta_k$ on  $L_k^t$  according to the formula $B = \sum_s
\beta_s B_s^t$,  set $d\sigma = a d\beta_1 ...  d\beta_k$,  where  $a$
does not depend on $t$.

{\bf Lemma  5.6.}  {\it  The  inner product $<\cdot,\cdot>_{L_k^t}$ is
invariant under time evolution:
$$
<\Psi_t,\Psi_t>_{L_k^t} = <\Psi_0,\Psi_0>_{L_k}.
$$
}

{\bf Proof.} By definition, one has
$$
<\Psi_t,\Psi_t>_{L_k^t} = a  \int  d\beta  (\Psi_t,  U[\sum_s  \beta_s
B_s^t] \Psi_t) = a \int d\beta (\Psi_0,  U_t^+ U[\sum_s \beta_s B_s^T]
U_t\Psi_0).
$$
Eq.\r{w23b} implies that
$$
U[\sum_s \beta_s B_s^t] = \exp \sum_s \int
\beta_s \it d{\bf x} (A_t^+({\bf
x}) B_s({\bf x}) - A_t^-({\bf x}) B_s^*({\bf x})).
$$
Making use of the relation
$$
U_t^+ A_t^{\pm} ({\bf x}) U_t = A^{\pm}({\bf x}),
$$
we obtain statement of lemma 5.6.

It follows  from lemma 5.6 that operator $U_t$ takes equivalent states
to equivalent. Therefore, it can be reduced to the factorspace
${\cal F}_{[k/2+1]}/\sim$.  Since it is unitary, it can be extended to
${\cal F}(L_k)$.

\subsection{Definition of a  composed  semiclassical  state  and  its
symmetry transformation}

Let us formulate a definition of a composed semiclassical state.

Let $\{X(\alpha), \alpha \in {\Lambda}^k \}$
be     a     smooth    $k$-dimensional    manifold
in   the   extended
phase space  $\cal  X$  with  measure $d\Sigma$ such that an isotropic
condition \r{1.21}
$$
\omega_{X(\alpha)} [\frac{\partial X(\alpha)}{\partial \alpha_a}] = 0.
$$
is satisfied.
It folows from commutation relations \r{1.15a} that
\beb
[\Omega_X [\frac{\partial X}{\partial \alpha_a};
\Omega_X [\frac{\partial X}{\partial \alpha_b}] =
i\left(
\frac{\partial \omega_i(X(\alpha))}{\partial \alpha_b}
\frac{\partial X_i}{\partial \alpha_a} -
\frac{\partial \omega_i(X(\alpha))}{\partial \alpha_a}
\frac{\partial X_i}{\partial \alpha_b}
\right)
= i
\left(
\frac{\partial}{\partial\alpha_b} (\omega_X            [\frac{\partial
X}{\partial\alpha_a}])
-
\frac{\partial}{\partial\alpha_a} (\omega_X            [\frac{\partial
X}{\partial\alpha_b}]
\right) = 0,
\eeb
so that
$$
[A^+B_a - A^-B_a^*;
A^+B_b - A^-B_b^*] = 0,
$$
where $B_a$ have the form \r{5.2}. Therefore,
$$
Im (B_a, B_b) = 0.
$$

Define an isotropic plane $L_k(\alpha) \equiv L_k(\alpha: \Lambda^k)$
as $span\{B_1,...,B_k\}$. It does not depend on the particular
choice  of coordinates $\alpha_1,...,\alpha_k$.
Introduce the following measure $d\sigma(\alpha)$ on $L_k(\alpha)$:
\be
d\sigma(\alpha) = \frac{D\Sigma(\alpha)}{D\alpha}(\alpha) d\beta_1 ...
d\beta_k,
\l{w25}
\ee
where $\beta_1$, ..., $\beta_k$ are coordinates on $L_k(\alpha)$ which
are determined as $B=\sum_s \beta_s B_s$.

Definition \r{w25}  is invariant under change of coordinates.  Namely,
let $(\alpha_1',...,\alpha_k')$ be another set  of  local  coordinates
chosen instead of $(\alpha_1,...,\alpha_k)$. Then
$$
B_l' = \sum_{s=1}^k \frac{\partial\alpha_s}{\partial\alpha_l'} B_s,
$$
so that  property  $\sum_l \beta_l' B_l' = \sum_s \beta_s B_s$ implies
that coordinate sets $\beta$ and $\beta'$ should be related as follows:
$$
\beta_s = \sum_l \frac{\partial\alpha_s}{\partial\alpha_l'} \beta_l'.
$$
Therefore, for the choice of coordinates $\alpha'$ one has
$$
d\sigma' = \frac{D\Sigma}{D\alpha'} d\beta_1'... d\beta_k' =
\frac{D\Sigma}{D\alpha'} \frac{D\alpha'}{D\alpha}     d\beta_1     ...
d\beta_k = d\sigma.
$$
The invariance property is checked.

Introduce the vector (Hilbert) bundle  $\pi_{\Lambda^k}$  as  follows.
The base  of  the  bundle  is the isotropic manifold $\Lambda^k$.  The
fibre that corresponds to the point $\alpha\in  \Lambda^k$  is  ${\cal
H}_{\alpha} = {\cal F}(L_k(\alpha))$.  Composed semiclassical states are
introduced as sections of bundle $\pi_{\Lambda^k}$.

{\bf Definition 5.2.} {\it A composed semiclassical state is a set  of
isotropic manifold   $\Lambda^k$   and   section  $Z$  of  the  bundle
$\pi_{\Lambda^k}$, such that the inner product
$$
<(\Lambda^k,Z), (\Lambda^k,Z)> = \int_{\Lambda^k} d\Sigma  (Z(\alpha),
Z(\alpha))_{{\cal F}(L_k(\alpha))}
$$
converges.
}

Group transformation   of   isotropic   manifold    $\Lambda^k    =
\{X(\alpha)\}$ is determined as
$$
u_{g} \{X(\alpha)\} = \{ u_{g} X(\alpha)\}.
$$
Section $\{Z(\alpha)\}$  is  transformed as follows.  Let $Z(\alpha) =
[Y(\alpha)]$. define   $U_{g}   Z(\alpha)   =   [U_{g}
Y(\alpha)]$. This  definition  is  correct  because  of the results of
previous subsubsection, provided that
\be
L_k(\alpha: u_{g}       \Lambda^k)       =       U_{g}
L_k(\alpha:\Lambda^k).
\l{w26}
\ee
It is sufficient to prove property \r{w26} for
the case $g = G_B(t)$. One should check that eq.\r{w23a}
\be
i\dot{B}_s^t = {\cal H}^{+-}(B: u_{g_B(t)}X) B_s^t
+ {\cal H}^{++}(B: u_{g_B(t)}X) (B_s^t)^*
\l{w27}
\ee
is satisfied for
$$
B_s^t = \varphi_{u_{g_B(t)}X} [\frac{\partial (u_{g_B(t)}X)
}{\partial\alpha_s}].
$$
However, system \r{w27} is a direct corollary of property F3.

Thus, the   composed   semiclassical   states   and   their  group
transformations are introduced.

\section{Conclusions}

Essential properties   of   the   semiclassical   Maslov   complex-WKB
approximation for  quantum  mechanics  -  a bundle structure of set of
semiclassical wave packets and their behavior under  small  variations
of classical  variables - are considered as a framework of an abstract
semiclassical mechanics. QFT models in the weak-coupling approximation
may be viewed as examples of abstract semiclassical systems.

Symmetry properties   of   semiclassical   systems   are   written  in
infinitesimal form.  Algebraic conditions \r{x4},  \r{x6} and \r{3.3},
\r{3.4} are  obtained  as infinitesimal analogs of semiclassical group
properties \r{1.5a},  \r{x0},  \r{1.23}, \r{1.24}. Condition \r{x6} is
very important  since  there  are  quantum  anomalies  in  QFT-models:
symmetry properties  may  be  violated   in   1-loope   approximation.
Therefore, satisfaction of relation \r{x6} means absense of anomalies.

It is  remarkable that the fact of commutativity of the left-hand side
of eq.\r{x6} with all operators $\Omega_X[\delta X]$ is a corollary of
eq.\r{3.4} and  classical  symmetry properties.  Thus,  one can expect
that identity \r{x6} is violated in quantum anomaly case in such a way
that its  right-hand  side  becomes  a  nontrivial  $c$-number (maybe,
$X$-dependent) quantity.

Sufficient conditions for constructing operators $U_g$ are  presented.
For the  case  of  $X$-independent  generators  $H(A:X)$,  they may be
viewed as an alternative for  known  conditions  of  integrability  of
Lie-algebra representations.

The obtained  properties  F1-F3  can be explicitly checked in proof of
Poincare invariance  of   hamiltonian   semiclassical   field   theory
\c{Shvedov}.

The semiclassical  Maslov  theory of Lagrangian manifolds with complex
germ (including WKB-method) may be also generalized to the case of the
abstract semiclassical  mechanics.  The  composed semiclassical states
are viewed  as  surfaces  on  the   semiclassical   bundle.   Symmetry
properties remain valid for the composed states as well.

This work  was supported by the Russian Foundation for Basic Research,
projects 99-01-01198 and 01-01-06251.


\newpage

\begin{thebibliography}{99}
\i{M1} V.P.Maslov, "Operational Methods", Moscow, Nauka, 1973;
English translation: Moscow, Mir Publishers, 1976.
\i{M2} V.P.Maslov, "The Complex-WKB Method for Nonlinear Equations",
Moscow, Nauka, 1977.
\i{J} R.Jackiw, {\it Rev.Mod.Phys.} {\bf 49} (1977), 681.
\i{DHN} R.Dashen, B.Hasslasher, A.Neveu, {\it Phys. Rev.} {\bf  D10}
(1974), 4114
\i{R} R.Rajaraman,  "Solitons  and  Instantons.  An  Introducteion   to
solitons     and     instantons     in    quantum    field    theory",
North-Holland,Amsterdam, Netherlands, 1982.
\i{J2} J. Coldstone, R.Jackiw, {\it Phys.Rev.} {\bf D11} (1975), 1486
\i{GMM} A.A. Grib, S.G. Mamaev, V.M. Mostepanenko,
"Vacuum Quantum Effects in Strong  Fields", Atomizdat, Moscow,
1988; Friedmann Laboratory Publishing, St. Petersburg 1994.
\i{BD} N.D. Birrell, P.C.W. Davies,
"Quantum Fields in Curved Space",
Cambridge, UK: Univ. Pr. , 1982.
\i{B1} D.Boyanovsky, H.J. de Vega and R.Holman,
{\it Phys. Rev.} {\bf  D49} (1994), 2769.
\i{B2} D.Boyanovsky, H.J. de Vega, R.Holman, D.S.Lee and A.Singh,
{\it Phys. Rev.} {\bf  D51} (1995), 4419.
\i{B11} Ju.Baacke, K.Heitmann and C.P\"atzold,
{\it Phys. Rev.} {\bf D55} (1997), 2320.
\i{B12} Ju.Baacke, K.Heitmann and C.P\"atzold,
{\it Phys. Rev.} {\bf  D56} (1997), 6556.
\i{HF1} F.Cooper, E.Mottola, {\it Phys.Rev.} {\bf D36} (1987), 3114
\i{HF2} S.-Y.Pi, M.Samiullah, {\it Phys. Rev.} {\bf D36} (1987), 3128
\i{G1} R.Jackiw and A.Kerman, {\it Phys.Lett.} {\bf A71} (1979), 158.
\i{G2} F.Cooper, S.-Y.Pi and P.Stancioff,
{\it  Phys. Rev.} {\bf D34} (1986), 3831.
\i{G3} O.Eboli, R.Jackiw and S.-Y.Pi,
{\it Phys. Rev.} {\bf D37} (1988), 3557.
\i{G4} O.Eboli,  S.-Y.Pi,  M.Samiullah,  {\it  Ann.  Phys.} {\bf 193}
(1989), 102.
\i{H} K. Hepp, "Theorie de la renormalisation",
Springer-Verlag, 1969.
\i{GJ} J. Glimm and A. Jaffe,
{\it "Boson quantum field models".}
In ''London 1971, Mathematics Of Contemporary Physics'', London 1972,
pp. 77-143.
\i{Ar1} I.Ya.Arefieva, {\it Teor.Mat.Fiz.} {\bf 14} (1973) 3.
\i{Ar2} I.Ya.Arefieva, {\it Teor.Mat.Fiz.} {\bf 15} (1973) 207.
\i{BS}
N.N. Bogoliubov, D.V. Shirkov, "Introduction to the
Theory of Quantized Fields", N.-Y.,Interscience Publishers, 1959.
\i{SF} A.A.Slavnov,  L.D.Faddeev,  "Introduction to the quantum theory
of gauge fields", Moscow, Nauka, 1988.
\i{A2} R.F.Streater,  A.S.  Wightman "PCT, spin and statistics and all
that", N.Y., Benjamin, 1964.
\i{A3}
N.N.Bogoliubov, A.A.Logunov, A.I.Oksak, I.T.Todorov, "General 
principles of Quantum  Field  Theory",  Moscow,  Nauka,  1987;  Kluwer
Academic Publishers, 1990.
\i{Maslov1} V.P.Maslov, "Perturbation Theory and Asymptotic Methods",
Moscow, Moscow University Press, 1965.
\i{MS3} V.P.Maslov,  O.Yu.Shvedov,  "The  Complex   Germ   Method   in
Many-Particle Problem  and  Quantum Field Theory",  Moscow,  Editorial
URSS, 2000.
\i{Shv1} O.Yu.Shvedov,  {\it Matematicheskie zametki} {\bf 65}  (1999)
437
\i{Shv2} O.Yu.Shvedov,  {\it Matematicheskii sborik} {\bf 190}  (1999)
N10, 123.
\i{P} M.M.Popov, {\it Zapiski Nauchnogo Seminara LOMI},{\bf 104}
(1981) 195.
\i{K} M.V.Karasev, {\it Zapiski Nauchnogo Seminara LOMI},{\bf 172}
(1989) 41.
\i{KV} M.V.Karasev, Yu.M.Vorobiev, preprint ITP-90-85E, Kiev, 1990.
\i{MS4} V.P.Maslov, O.Yu.Shvedov {\it Teor. Mat.Fiz.} {\bf 104} (1995)
479.
\i{MS-FT} V.P.Maslov,  O.Yu.Shvedov  {\it  Teor.  Mat.Fiz.}  {\bf 114}
(1988) 233.
\bibitem{BR} A.Barut and R.Raczka,  Theory of group representations
and applications, Warszawa, Pol.sci.publ., 1977.
\bibitem{N} E. Nelson, {\it Ann. Math.} {\bf 70}, 572  (1959)
\bibitem{FS1} M.Flato, J.Simon, H.Snellman and D.Sternheimer,
{\it Ann. Scient. de l'Ecole Norm. Sup.} {\bf 5}, 423 (1972).
\bibitem{FS2}
J.Simon, {\it Comm. Math. Phys.} {\bf 28}, 39 (1972).
\bibitem{FS3}
M.Flato and J.Simon, {\it J. Funct. Anal.} {\bf 13}, 268 (1973).
\i{KF} A.N.Kolmogorov, S.V.Fomin,  "Elements of Functions  Theory  and
Functional Analysis", Moscow, Nauka, 1989.
\i{Pontriagin} L.S.Pontriagin, "Continous groups", Moscow, Nauka, 1973.
\i{Ber} F.A.Berezin,  "The Method  of  Second  Quantization",  Moscow,
Nauka, 1965; N.Y.1996.
\i{Shvedov} O.Yu.Shvedov, hep-th/0109142.
\i{MS-RJMP} V.P.Maslov, O.Yu.Shvedov, {\it Russian J. Math.Phys.}
4 (1996) 173.
\i{KA} L.V.Kantorovich, G.P.Akilov, "Functional Analysis",
Moscow, Nauka, 1984.
\i{Marolf} A.Ashtekar, J.Lewandowski, D.Marolf, J.Mourao and T.Thiemann,
{\it J. Math. Phys.} 36 (1995) 6456.
\end{thebibliography}
\end{document}